\documentclass[10pt, conference, letterpaper]{IEEEtran}

\usepackage[cmex10]{amsmath}
\usepackage[tight,footnotesize]{subfigure}
\usepackage{url}
\usepackage[tight,footnotesize]{subfigure}
\usepackage{graphicx}
\usepackage{graphics}
\usepackage{amssymb}
\usepackage{amsmath}
\usepackage{amsthm}
\usepackage{color}
\theoremstyle{definition}

\usepackage{tikz}

\usepackage[hang,small,bf]{caption}

\newcommand{\figref}[1]{\figurename~\ref{#1}}

\newcommand{\nop}[1]{}

\usepackage{lipsum}
\usepackage{xcolor}
\usepackage{soul}
\sethlcolor{black}
\makeatletter
\newif\if@blind
\@blindtrue %use \@blindfalse on final version

\if@blind \sethlcolor{black}\else
   
\fi
\begin{document}
%
% paper title
% can use linebreaks \\ within to get better formatting as desired

%%%% different title candidates%%%
%\title{Experimental Study of Cross-layer Effects on Multiple Performance Metrics for 802.15.4 link}

%\title{How to Optimize Your Stack Parameters: Insights of an Experimental Study for Communication over 802.15.4 Link}
%\title{How to Optimize Your Stack Parameters: Insights of a Cross-layer Experimental Study for Wireless Link in Sensor Networks}

%\title{Towards Performance Optimization: A Cross-layer Experimental Study for WSN Link Communication}
%\title{Towards understanding complex performance behaviors: A cross layer measurement study for wireless link in WSN}
%\title{A cross layer measurement study towards understanding wireless link performance in sensor networks}

%\title{A Cross-layer Measurement Study to Improve Wireless Link Performance in Sensor Networks}

%\title{A Cross-layer Measurement Study Towards Understanding Wireless Link Performance in WSN}
\title{An Experimental Study Towards Understanding Data Delivery Performance Over a WSN Link}

% author names and affiliations
% use a multiple column layout for up to three different
% affiliations			
\author{
    \IEEEauthorblockN{Songwei Fu\IEEEauthorrefmark{1}, Yan Zhang\IEEEauthorrefmark{2}, Yuming Jiang\IEEEauthorrefmark{2}, Chia-Yen Shih\IEEEauthorrefmark{1}, Pedro J\'{o}se Marron\IEEEauthorrefmark{1}}
    \IEEEauthorblockA{\IEEEauthorrefmark{1}Networked Embedded Systems, University of Duisburg-Essen, Germany
    \\\{songwei.fu, chia-yen.shih, pjmarron\}@uni-due.de}
    \IEEEauthorblockA{\IEEEauthorrefmark{2}Department of Telematics, Norwegian University of Science and Technology(NTNU), Norway
    \\\{yanzhang, ymjiang\}@item.ntnu.no}
}

%\author{}

% make the title area
\maketitle

\begin{abstract}
% IEEEtran.cls defaults to using nonbold math in the Abstract.
% This preserves the distinction between vectors and scalars. However,
% if the conference you are submitting to favors bold math in the abstract,
% then you can use LaTeX's standard command \boldmath at the very start
% of the abstract to achieve this. Many IEEE journals/conferences frown on
% math in the abstract anyway.

The performance of a wireless sensor network (WSN) depends fundamentally on how its various parameters are configured under different link quality conditions. Surprisingly, even though WSNs have been extensively researched, there still lacks an in-depth understanding on how parameter configurations affect, in particular jointly, the performance under different link quality conditions. To fill the gap, this paper presents an extensive experimental study on the performance of a wireless sensor network link, where measurement data of more than 200 million packets were collected. Different from existing work, we consider major parameters from different layers {\em together} and measure their {\em joint effects} on key performance metrics under {\em an extensive set of parameter configurations}. Based on the large amount of measurement data, we investigate the impacts of these parameters and their joint configurations on the performance, introduce empirical models to theoretically reason the impacts, discuss implications of the obtained results, and suggest practical guidelines for parameter configurations. Through these, a comprehensive overview of parameter configuration impacts on the performance is provided, which provides new insights on wireless link performance in WSNs. %under different channel conditions 
%\keywords{IEEE 802.15.4, cross layer effects, experimental study} % NOT required for Proceedings
\end{abstract}

\section{Introduction}\label{sec:intro}

Many wireless sensor network (WSN) applications have quality-of-service (QoS) requirements on the underlying WSNs to achieve desired application performance. Example application scenarios include structure-health monitoring\cite{5211924, Chipara:2010:RCM:1869983.1869999}, audio/video transmission for surveillance \cite{4653057}, health monitoring \cite{4379685} and bulk data collection \cite{Kim:2007:FRB:1322263.1322296, Werner-Allen:2006:FYV:1298455.1298491}, where, in addition to providing a suitable throughput, the network also needs to cope with delay constraints while considering the limitation of energy consumption. 

The performance of a wireless sensor network depends fundamentally on how its various parameters are configured under different link quality conditions. To gain an in-depth understanding of the parameter configuration impact on the performance, conducting experiments is a crucial step. Several such studies have been reported in recent years including cross-layer analysis for delay \cite{Ferrari:2007:WSN:1283620.1283661} \cite{6424813} or packet loss rate \cite{Zhao:2003:UPD:958491.958493} \cite{1683512} \cite{6662477} \cite{5935312}, and the impact of packet length \cite{5462063},  transmission power \cite{Lin:2006:AAT:1182807.1182830}, or MAC layer parameters \cite{Son06experimentalstudy} \cite{6680307} \cite{Anastasi:2009:MUP:1641804.1641839} \cite{Zimmerling:2012:PRP:2185677.2185730} on the performance. However, these studies only consider one single chosen parameter or single layer and try to optimize the system performance via the chosen parameter, and hence only reveal partly the overall picture. As a consequence, if not used properly, their results may lead to inaccurate modeling of the system and hence cannot be used as the best suitable guidelines for parameter configurations in WSNs.

To fill the gap, we report in this paper an extensive experimental study on the performance of a wireless sensor network link adopting IEEE 802.15.4. Major related parameters from different layers are {\em considered together}, which include {\em transmission power at the physical layer (PHY layer), maximal number of retransmissions, holding time before each retransmission, and maximal queue size at the MAC layer, and packet inter-arrival time and packet payload size at the application layer, under different distances between the sensors}. Their {\em joint effects} on key performance metrics, including {\em energy efficiency and QoS metrics -- delay, goodput and packet loss rate}, under {\em an extensive set of parameter configurations} are measured. Altogether, close to {\bf 50 thousand} parameter configurations were experimented and measurement data of more than 200 million packets were collected over a period of 6 months. (This dataset will be made publicly available.) To the best of our knowledge, this is the first study at this detailed level. %(Cf. Sec. \ref{sec:setup})

Based on the measurement data, we investigate what and how these stack parameters jointly impact QoS metrics and energy consumption, and what and how parameters may be adjusted to improve the performance. The specific contributions of the paper are as follows, which are novel and provide a comprehensive overview of parameter configuration impacts on wireless link performance in WSNs:
\begin{itemize}
  \item The impacts of the various parameters and their joint configurations on the performance are detailedly investigated (Cf. Sec. \ref{sec:radio-channel} and Sec. \ref{sec:analysis}).  
  \item Several empirical models are introduced to theoretically reason the impacts (Cf. Eqs. (\ref{equ:ro}), (\ref{avg_servicetime1}) and (\ref{avg_servicetime2})).
  \item The implications of the obtained results are discussed, based on which practical guidelines for parameter configurations are suggested (Cf. Sec. \ref{sec:analysis})
  \item The impacts of link quality condition on parameter configurations and the considerations of using the corresponding results are investigated (Cf. Sec. \ref{sec:dis}).  
  \item The trade-offs of important stack parameters on the performance metrics and energy consumption are discussed, helpful for stack parameter optimization when multiple performance metrics are considered together as system requirements (Cf. Sec. \ref{sec:dis}).
\end{itemize}

The rest is organized as follows. The experiment setup is introduced in Sec. \ref{sec:setup}. In Sec. \ref{sec:radio-channel}, we present results mainly related to the physical layer, including channel characteristics and packet error rate. In Sec. \ref{sec:analysis}, results for energy consumption and QoS related performance metrics including delay, goodput and packet loss rate are given. In Sec. \ref{sec:dis}, additional discussion on the implication of the results is made. Finally, we conclude the paper in Sec. \ref{sec:conclusion}.

\section{Experimental Methodology}\label{sec:setup}

\subsection{Experiment Setup}

To achieve the goal of this study, i.e, to obtain an in-depth understanding of wireless link performance in WSNs, we conducted an extensive set of experiments in an indoor office building environment. We employed a sender-receiver pair of TelosB motes, each equipped with a TI CC2420 radio using the IEEE 802.15.4 stack implementation in TinyOS. In each experiment, the sender sends packets to the receiver under a particular stack parameter configuration. {\em Despite the simplicity of the setup, the fundamental properties of IEEE 802.15.4 wireless link are revealed}. The whole experiment took more than 6 months between November 2012 and November 2013.  

\begin{figure}[ht!]
\centering
\includegraphics[width=1.0\columnwidth]{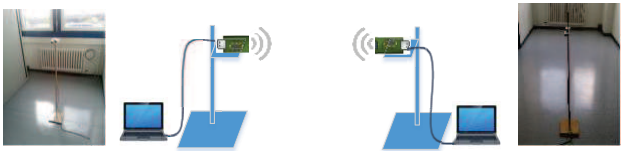}
\caption{Experiment setup}% change the name
\label{expFigure2}
\end{figure}

\begin{figure}[ht!]
\centering
\includegraphics[width=1.0\columnwidth]{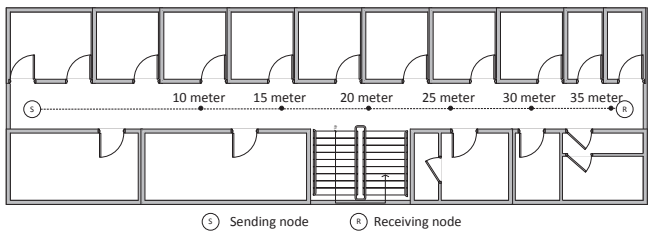}
\caption{Floor plan with positions of sensor nodes}% change the name
\label{expFigure1}
\end{figure}

Specifically the experiments were conducted in a long hallway of 2 meters by 40 meters. Each mote was fixed on a wooden stand of 0.7-meter high, and the antennas of the two motes were facing each other, as shown in~\figref{expFigure2}. For each experiment, we maintained a line-of-sight between the two motes at a specific distance, which varied for different experiments ranging from 10 meters to 35 meters. \figref{expFigure1} illustrates the building floor plan and the node positions. This hallway poses a particularly harsh wireless environment because of significant likelihood of multi-path reflection from the walls. In addition, university students and employees may walk in the hallway. Moreover, the hallway is in the range of several WiFi access points. 

%figure of corridor and floor plan

\subsection{Parameter Configuration}

The IEEE 802.15.4 standard \cite{4040993} specifies both Physical (PHY) and Medium Access Control (MAC) layer parameters. At the PHY layer, the radio CC2420 achieves a data rate of 250 kb/s using O-QPSK in the ISM band of 2.4 Ghz, and provides programmable \textit{transmission power} $P_{TX}$ of 8 levels.
At the MAC layer, IEEE 802.15.4 supports both beacon and beaconless modes. In our experiments, the beaconless mode with unslotted CSMA-CA for channel access was used. In addition, the ACK frame was enabled to allow MAC layer retransmission. Corresponding to this, two MAC parameters were considered in the experiment: \textit{maxRetries} and  \textit{retry delay} $D_{retry}$, which are the maximal number of (re-)transmissions to deliver a packet, and the waiting time for a new retransmission, respectively. A FIFO with tail-drop queue on top of the MAC layer was used to buffer application packets when they are waiting for (re-)transmission. The corresponding configurable parameter was the \textit{maximal queue size} $Q_{max}$. 

At the application layer, we adopted two common parameters: \textit{packet inter-arrival time} $T_{pit}$ and \textit{packet payload size} $l_D$ for generating traffic at different workload levels. 

Table~\ref{tab:ExpParameter} summarizes the aforementioned parameters, together with the distance $d$, and their values used in the experiments. In total, there are 8064 stack parameter configurations for each distance. With the distance, the number of experimented parameter configurations is close to {\bf 50 thousand}.
 
\begin{table}[hbt]
	\centering
			\caption{Stack Parameter Configurations}
		\begin{tabular}{|c|l|c|}
		\hline
		\textbf{Layer} & \textbf{Parameters} & \textbf{Parameter values} \\ 
		\hline
		\hline
		\textbf{Appl.} & $T_{pit}$: packet inter-arrival time (ms) & 10,15,20,25,30,35,40,50 \\
		%\hline			
		& $l_D$: packet payload size (bytes) & 20,35,50,65,80,95,110 \\
		\hline
		\textbf{MAC}	& $Q_{max}$: maximal queue size & 1,30,60 \\
		%\hline
	  & maxRetries: maximal \# of retires & 1,3,5 \\
		%\hline
		& $D_{retry}$: retry delay (ms) & 30,60 \\
		\hline
		\textbf{PHY}	& $P_{TX}$: transmission power level & 3,7,11,15,19,23,27,31 \\
		%\hline
		 & d: distance between nodes (meter) & 10,15,20,25,30,35 \\
		\hline
		\end{tabular}
	\label{tab:ExpParameter}
\end{table}

\subsection{Experiment Design and Information Collection}

For each distance and stack parameter configuration shown in Table~\ref{tab:ExpParameter}, the sending node continuously sent 300 packets using the same configuration. Both the sending and receiving nodes logged per-transmission information including received signal strength, time of receiving, actual transmission number, actual queue size, etc. We refer this process as one experiment run and each experiment was repeated 15 runs to acquire sufficient information for statistical analysis. On average, for each distance, experimenting all the stack parameter configurations took approximately one month. 

In the end, the transmission information for more than {\bf 200 million} packets was collected. Based on the collected information, we conduct analysis and discuss findings in Sec. \ref{sec:radio-channel} and Sec. \ref{sec:analysis} from the aspects of {\em radio channel characteristics}, {\em energy consumption}, and {\em QoS related performance metrics, including delay, goodput and packet loss rate}.
 
\section{Experimental Results: Channel Characteristics and Packet Error Rate}\label{sec:radio-channel}

In this section, we focus on channel characteristics and Packet Error Rate (PER). While the former is fundamental for understanding the experiment environment, the latter is an important step towards analyzing other performance metrics.

\subsection{Radio Channel Characteristics}

\subsubsection{Radio signal attenuation}
\label{sec:RSSIMeasurement}

\begin{figure*}[htb!]
\begin{minipage}[t]{0.33\linewidth}
\centering
\includegraphics[width=2.0in, height=1.5in]{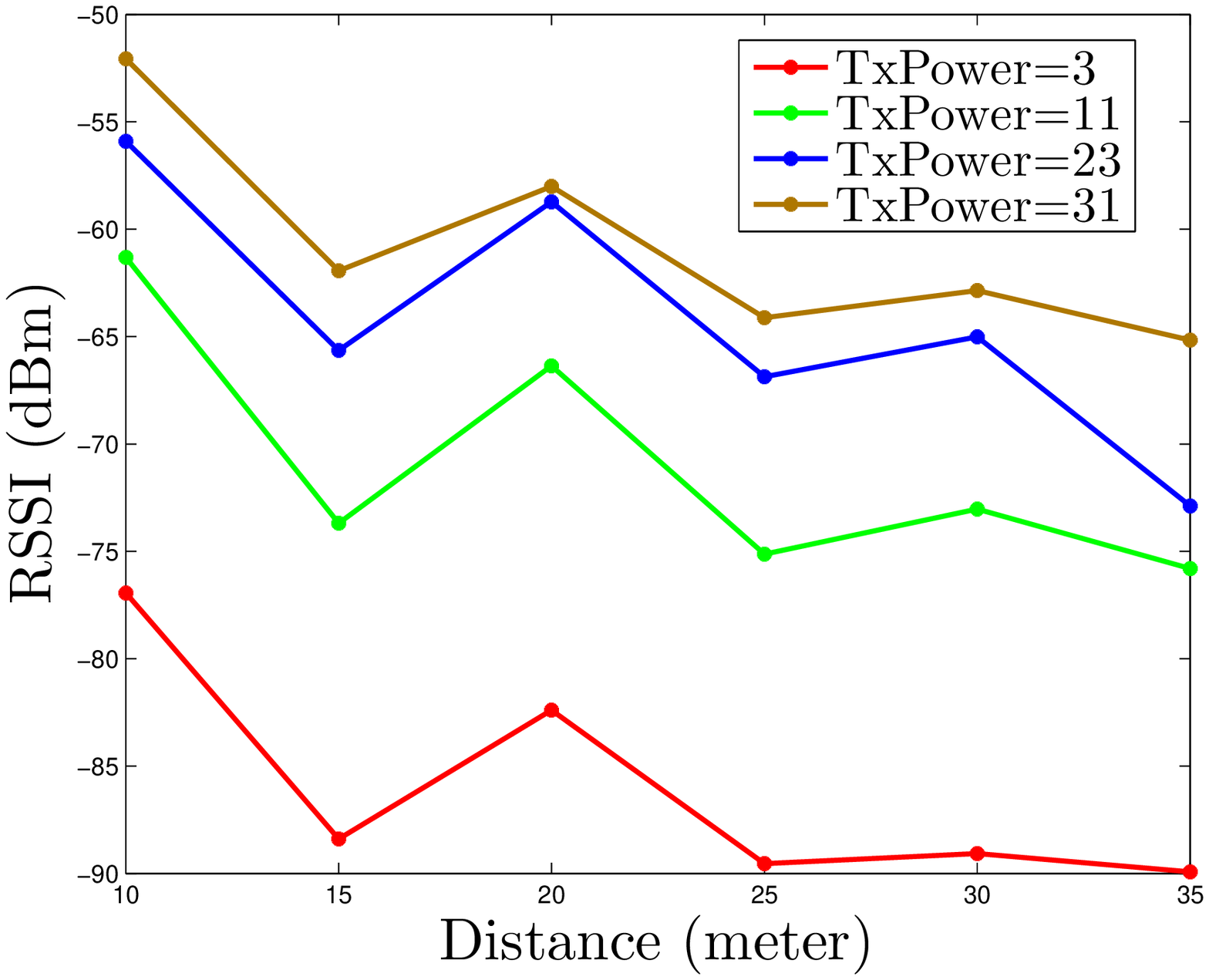}
\caption{Average RSSI vs. distance }
\label{avgrssi}
\end{minipage}%
\begin{minipage}[t]{0.33\linewidth}
\centering
\includegraphics[width=2.0in, height=1.5in]{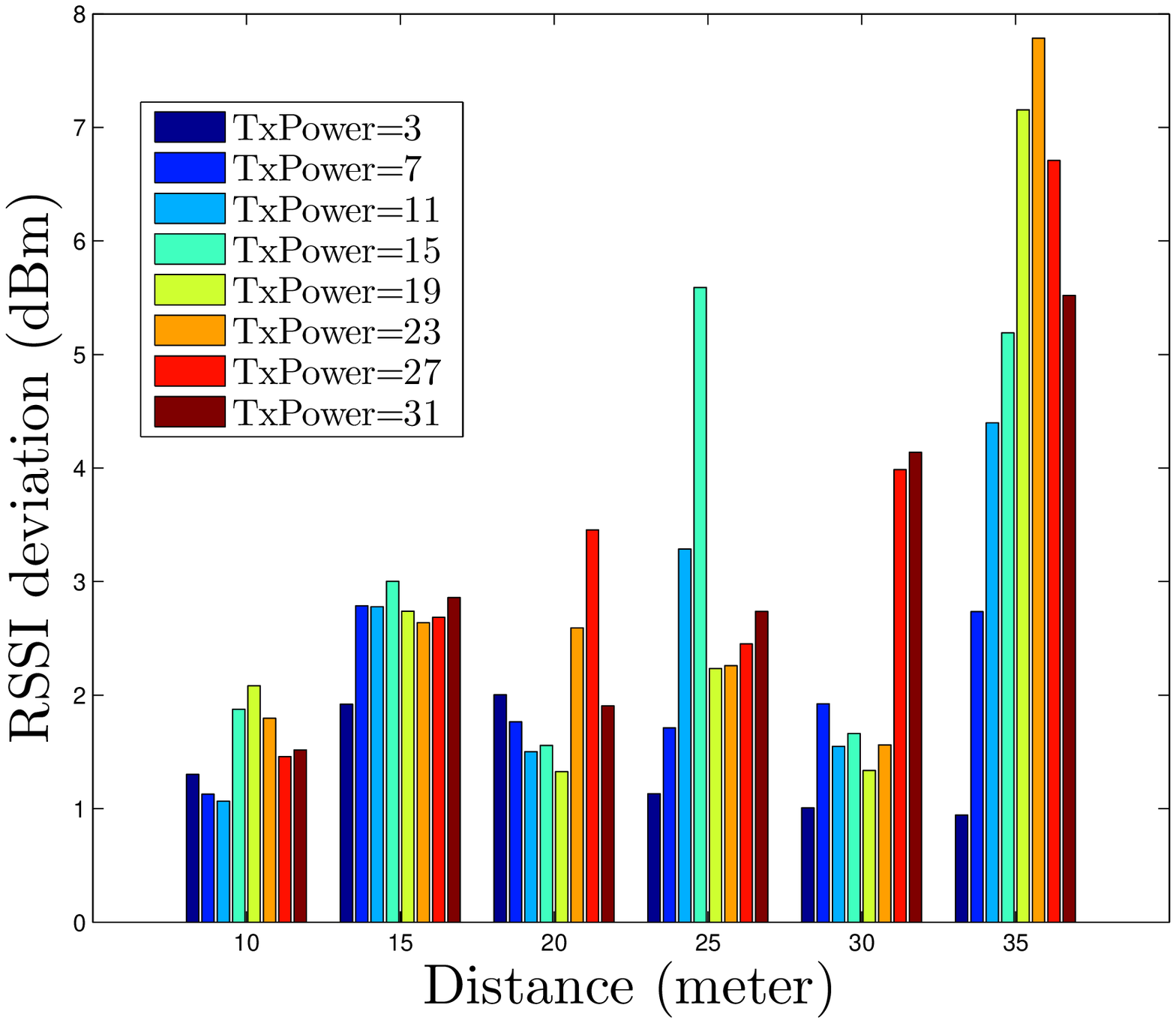}
\caption{Standard deviation of RSSI}
\label{SDrssi}
\end{minipage}%
\begin{minipage}[t]{0.33\linewidth}
\centering
\includegraphics[width=2.0in, height=1.5in]{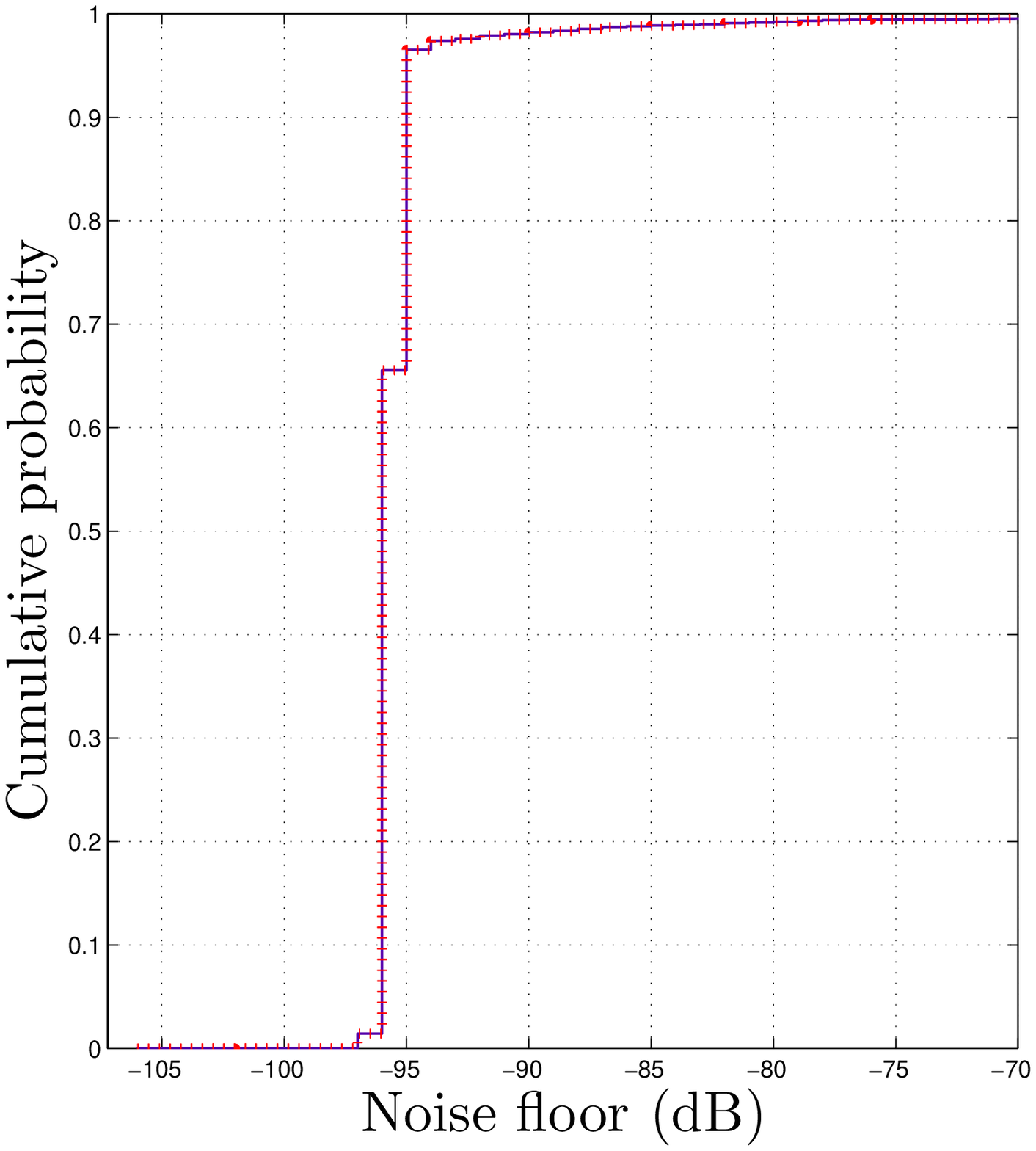}
\caption{Noise floor distribution }
\label{CDFnoise}
\end{minipage}%
\end{figure*}

To see how signal attenuates over distance in our experiment environment, we show the average RSSI (Received Signal Strength Indicator) for every transmission power levels $P_{TX}$ with increasing distances. \figref{avgrssi} shows that the average RSSI decreases approximately linearly as a function of the distance in logarithmic scale for all $P_{TX}$. This indicates that the path loss in our indoor office building environment can be well modeled with the log-normal shadowing model \cite{Rappaport:1996:WCP:525688}:
\begin{equation}\label{pathloss}
   \bar{PL}(d)[dB]=\bar{PL}(d_{0})+10n\log_{10}(\frac{d}{d_0})+\chi_\sigma(t)
\end{equation}
where $d_{0}=1m$ is a reference distance, $n$ is the path loss factor, $\chi_\sigma(t)$ is a zero-mean Gaussian random variable (shadowing effects) with standard deviation $\sigma$. Using Matlab curve fitting method, we obtain the parameters for our experiment environment as $\bar{PL}(d_{0})=32.01,n=2.19$ and $\sigma=3.2$.

\subsubsection{Radio signal fading}

Radio signal fading, resulted from the properties of the environment such as walls, is another important characteristic of a wireless channel. The variance of RSSI values reflects the signal fading effect. To investigate how the RSSI value varies with node positions and $P_{TX}$, we plot the RSSI deviation for each $P_{TX}$ level at different distances in ~\figref{SDrssi}. The figure shows that the RSSI varies under the same $P_{TX}$ and distance, and that there is no clear correlation between the RSSI variance, $P_{TX}$ and distance.

\subsubsection{Noise floor}
The third channel characteristic we investigate is the noise floor, which is the measured RSSI when the channel is idle. It is mainly contributed by the radio components and interfering signals. ~\figref{CDFnoise} shows the Cumulative Distribution Function (CDF) of approximately 24 million noise floor samples collected in the experiment. The figure indicates that {\em while the noise floor is generally stable (with average noise level of approximately -95.4dBm in our experiment environment), it cannot or should not be treated as a constant value}. 

\vspace{3mm}
In the rest of the paper, for ease of representation and as commonly used in the literature, we use signal-to-noise-ratio (SNR), defined to be RSSI/(noise floor), as the combined effect of the RSSI and the noise floor to represent the channel characteristics / quality condition.

\subsection{Packet Error Rate}
\label{sec:per}

Packet error rate (PER) is the ratio of the number of incorrectly received data packets to the total number of transferred packets, and PER can be calculated as: 
 \begin{equation}\label{equ:per}
   {PER} = \frac{ \textrm{\# of non-ACKed transmissions}}{\textrm{\# of total transmissions}}.
\end{equation}
PER is the immediate performance metric of the physical layer to the upper layer, affected by the physical layer condition or more specifically SNR and the packet payload size $l_D$.

\begin{figure*}
\begin{minipage}[t]{0.33\linewidth}
\centering
\includegraphics[width=2.2in, height=1.5in]{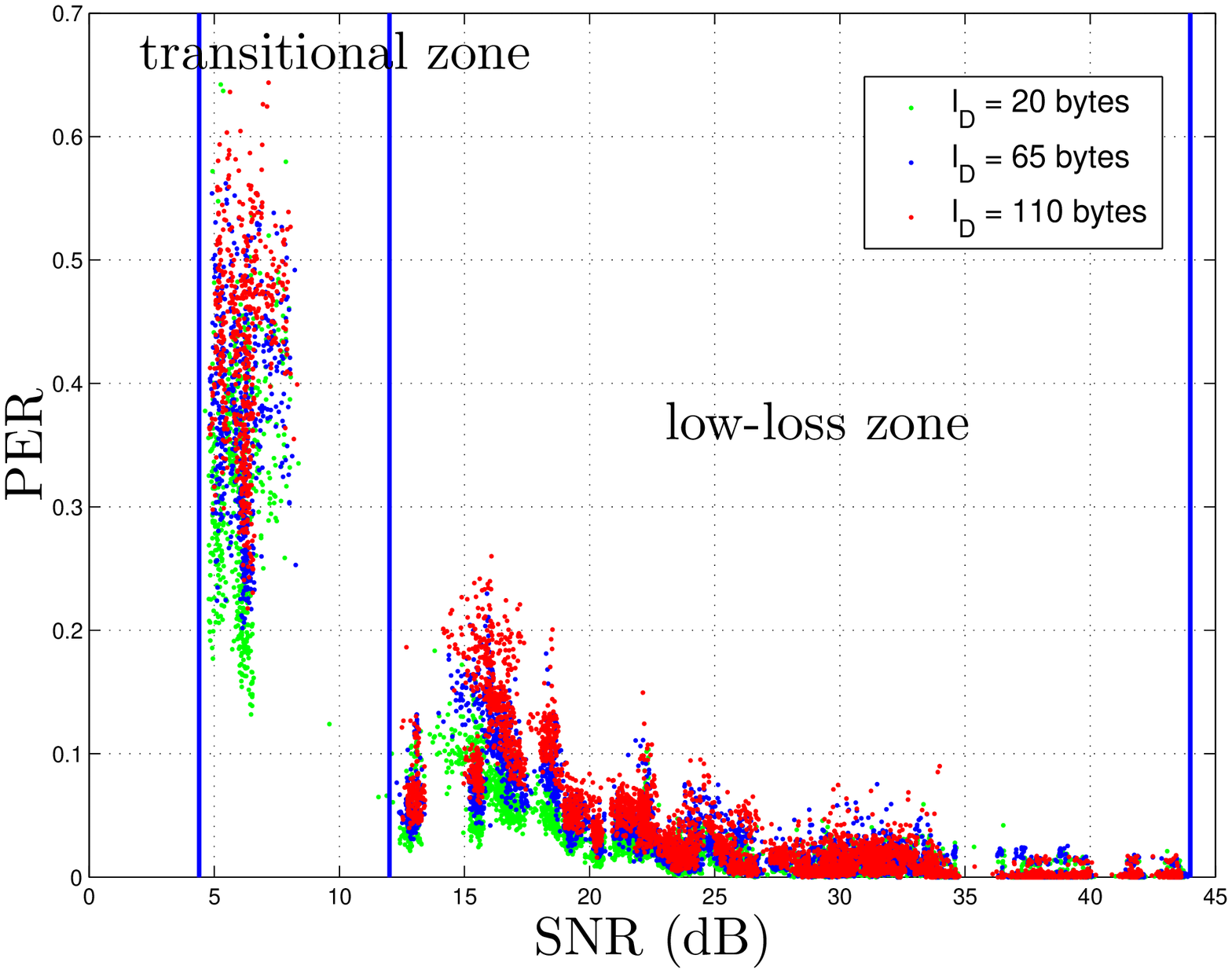}
\caption{Scattered PER values}
\label{PER1}
\end{minipage}%
\begin{minipage}[t]{0.33\linewidth}
\centering
\includegraphics[width=2.2in, height=1.5in]{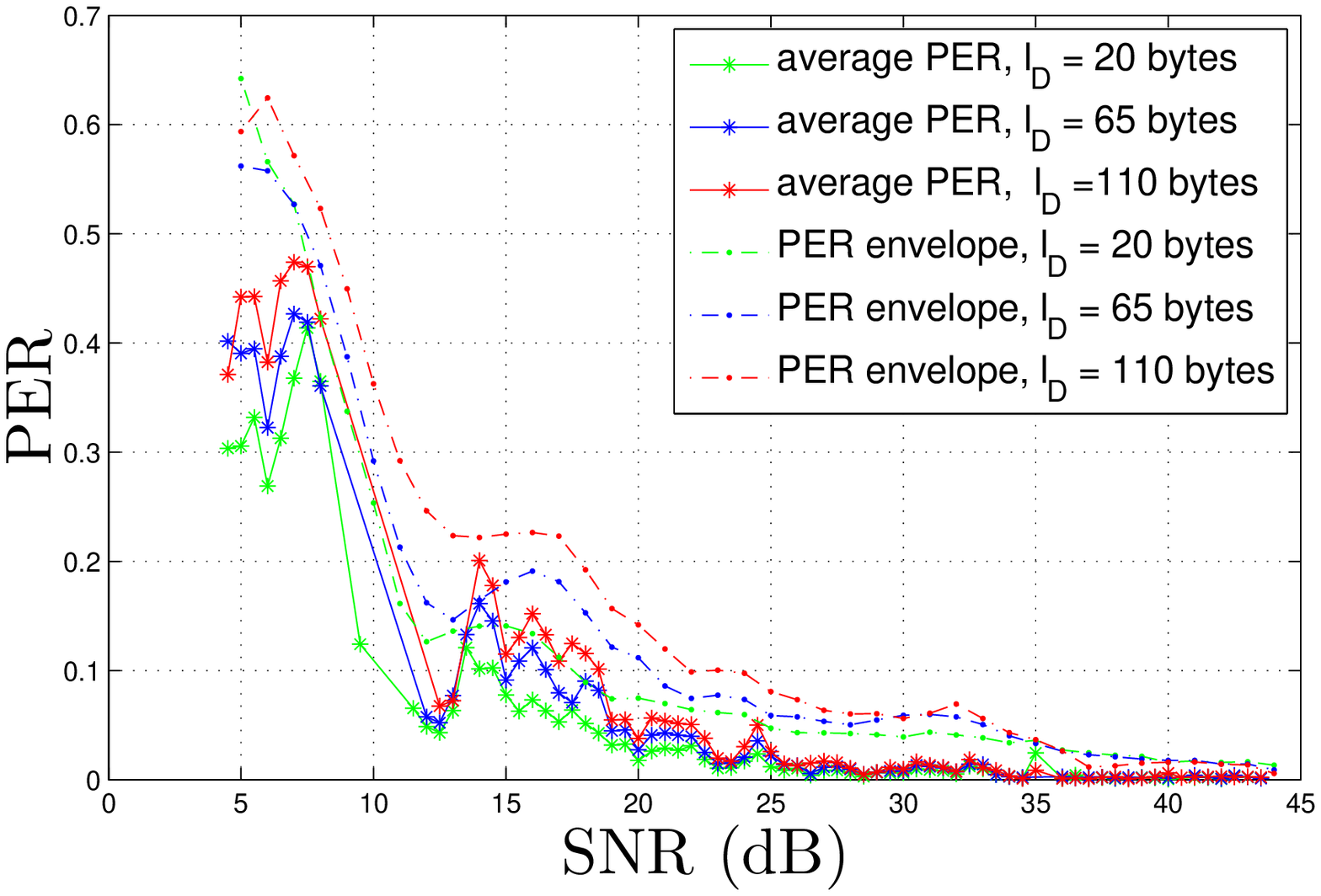}
\caption{Average PER and PER envelope}
\label{PER2}
\end{minipage}%
\begin{minipage}[t]{0.33\linewidth}
\centering
\includegraphics[width=2.5in, height=1.5in]{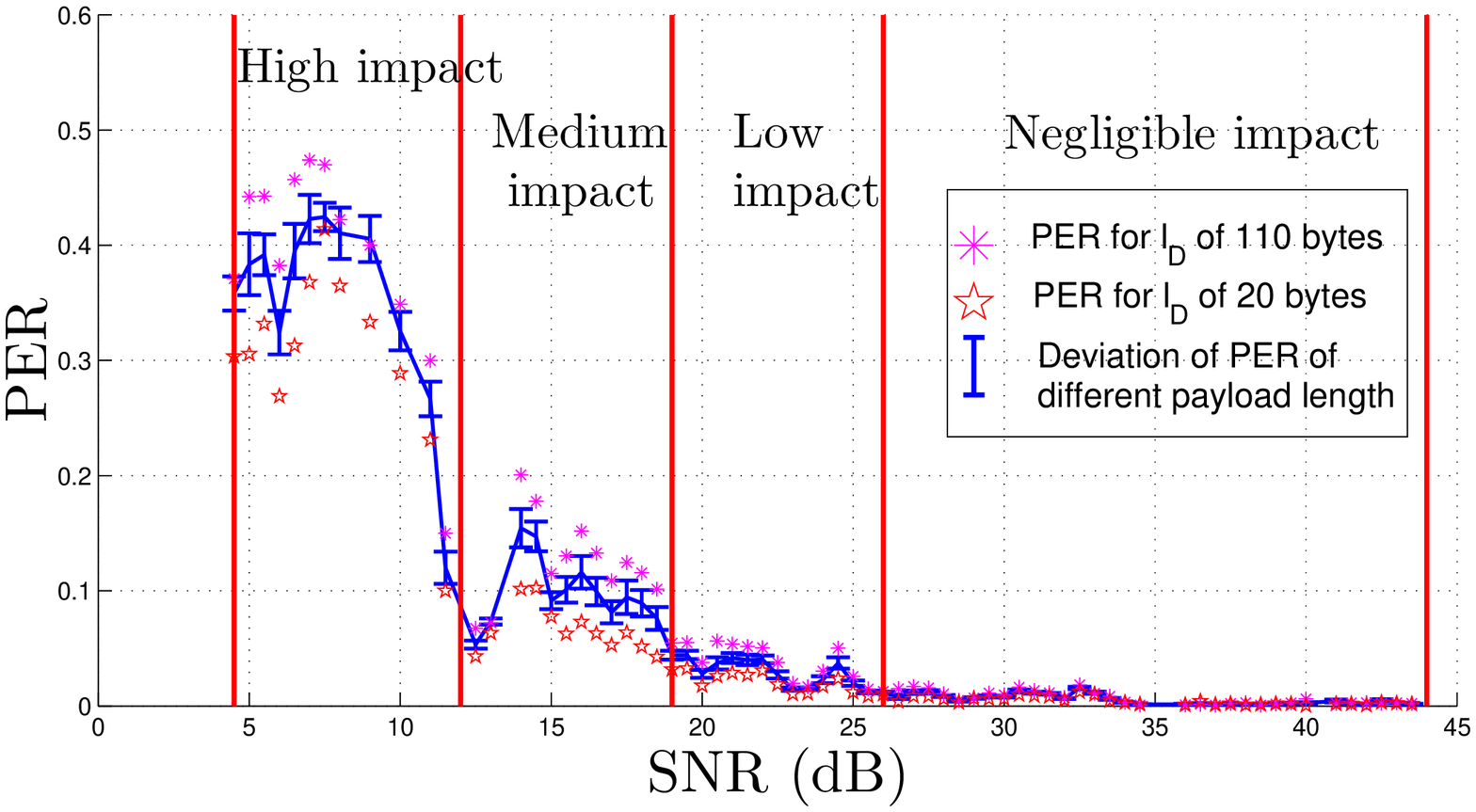}
\caption{The impact of payload size on PER}
\label{PER3}
\end{minipage}%
\end{figure*}

\subsubsection{Transitional zone and low-loss zone} 

To show how PER changes with SNR, the scattered PER values, and the average PER with envelope (the upper bound of PER) are plotted against SNR in~\figref{PER1} and~\figref{PER2} respectively. It can be observed that our PER measurements match with the existing PER-based classification of links \cite{1381954}, which is (1) a low-loss zone where the links observe very few packet losses and (2) an intermediate transitional zone where the link quality fluctuates between low and high losses. We do not have the link in the high-loss zone due to the limited space of the hallway. As illustrated in~\figref{PER1} and~\figref{PER2}, the cross-over SNR value from the transitional zone to the low-loss zone is approximately 12dB for our environment. In addition, it can be observed that the cross-over points shift to the right with the increase of payload size. More discussion on the effect of payload size is in the following.

\subsubsection{Impact of payload size on PER} 

To further examine how PER responds to payload size $l_D$, we compute the average PER for each $l_D$ at various SNR values, and the standard deviation of the average PER values of different $l_D$.  The results are presented in~\figref{PER3}. It can be observed that while the standard deviation generally deceases with SNR, it has similar values in certain SNR regions. The region from 4dB to 12dB has the highest standard deviation (0.03 - 0.04) and the region from 12dB to 19dB has the medium standard deviation (0.02 - 0.03). We refer these SNR regions as the high-impact zone and the medium-impact zone of payload size on PER respectively. Similarly, we define the low-impact zone and negligible-impact zone of $l_D$ on PER, as marked in~\figref{PER3}.

\vspace{3mm}
In brief, the above results indicate that {\em there is a tremendous decrease of PER if the SNR level crosses the border between the transitional zone and the low-loss zone} (12dB in our case). In addition, {\em the impact of payload size on PER tends to be small only after SNR crosses the border between the medium-impact and the low-impact zone} (19dB in our case). The distinction of the zones implies that the observation from parameter configurations under PER conditions in one zone cannot or should not be generalized for parameter configurations under PER conditions in other zones. %In other words, special care is needed when using results that do not reflect the whole picture. 

\section{Experimental Results: Energy Efficiency, Delay, Goodput and Packet Loss Rate}\label{sec:analysis}

The focus of this section is on the joint impact of (multi-layer) parameter configurations on energy efficiency and QoS performance -- delay, goodput and packet loss rate.

%%%%%%%%%%%%%%%%%%%%%%%%%%%%%%%%%%%%%%%%%%%%%% Energy efficiency %%%%%%%%%%%%%%%%%%%%%%%%%%%%%%%%%%%%%%%%%%%%%%%%%%%%%%%%%%%

\subsection{Energy Efficiency}
\label{sec:energy}

Energy efficiency is a crucial performance metric of WSN. The energy spent for transmitting a packet depends on the energy required for each transmission and the PER. Specifically, we define the energy consumption of transmitting per information bit $U_{eng}$ as follows. 
\begin{equation} \label{EnergyPerBit}
{U_{eng}} = \frac{{{E_{tx}} \cdot ({l_O} + {l_D})}}{{{l_D} \cdot (1-\mathrm{PER})}}
\end{equation} 
where $l_O$ and $l_D$ are the stack overhead size and packet payload size respectively. $PER$ is the average packet error rate for a given transmission power $P_{TX}$ and distance. $E_{tx}$ is the energy consumption for transmitting one bit and can be computed by $I_{tx}\cdot V_{DD}\cdot t_{tx}$. From the datasheet of CC2420, the current consumption $I_{tx}$ of each $P_{TX}$ is available. The supply voltage $V_{DD}$ is 1.8V, and the time of transmitting one bit $t_{tx}$ is $4\mu s$ at 250kpbs. 

\begin{figure}[htbp]
\centering
\includegraphics[width=1.0\columnwidth]{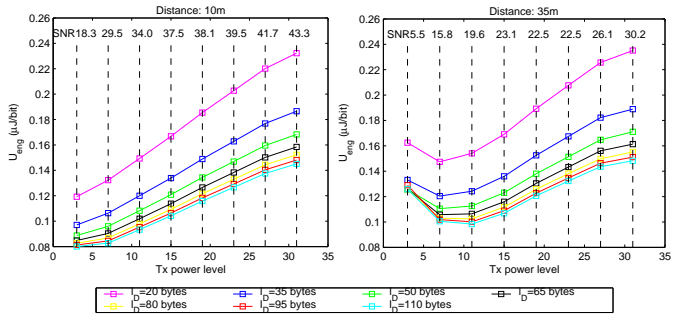}
\caption{$U_{eng}$ at distances 10m and 35m}
\label{Energy1}
\end{figure}

\textbf{Observation.} In order to understand the joint-impacts of $l_D$ and $P_{TX}$ on energy consumption, ~\figref{Energy1} plots $U_{eng}$ against $P_{TX}$ for different $l_D$ at the distances of 10m and 35m. $U_{eng}$ of other distances are not shown here because they have similar behaviors as at distance 10m. It can be observed that a larger $l_D$ generally leads to a better $U_{eng}$ except for $P_{TX} = 3$ at distance 35m. In addition and more interestingly, {\em the trend of $U_{eng}$ against $P_{TX}$ is different for 10m and for 35m}. Specifically, while $U_{eng}$ increases monotonically with $P_{TX}$ for 10m, it first decreases when $P_{TX}$ increases from 3 to 7 and then increases with $P_{TX}$ for 35m. 

\begin{figure}
\begin{minipage}[t]{0.5\columnwidth}
\centering
\includegraphics[width=1.9in, height=1.8in]{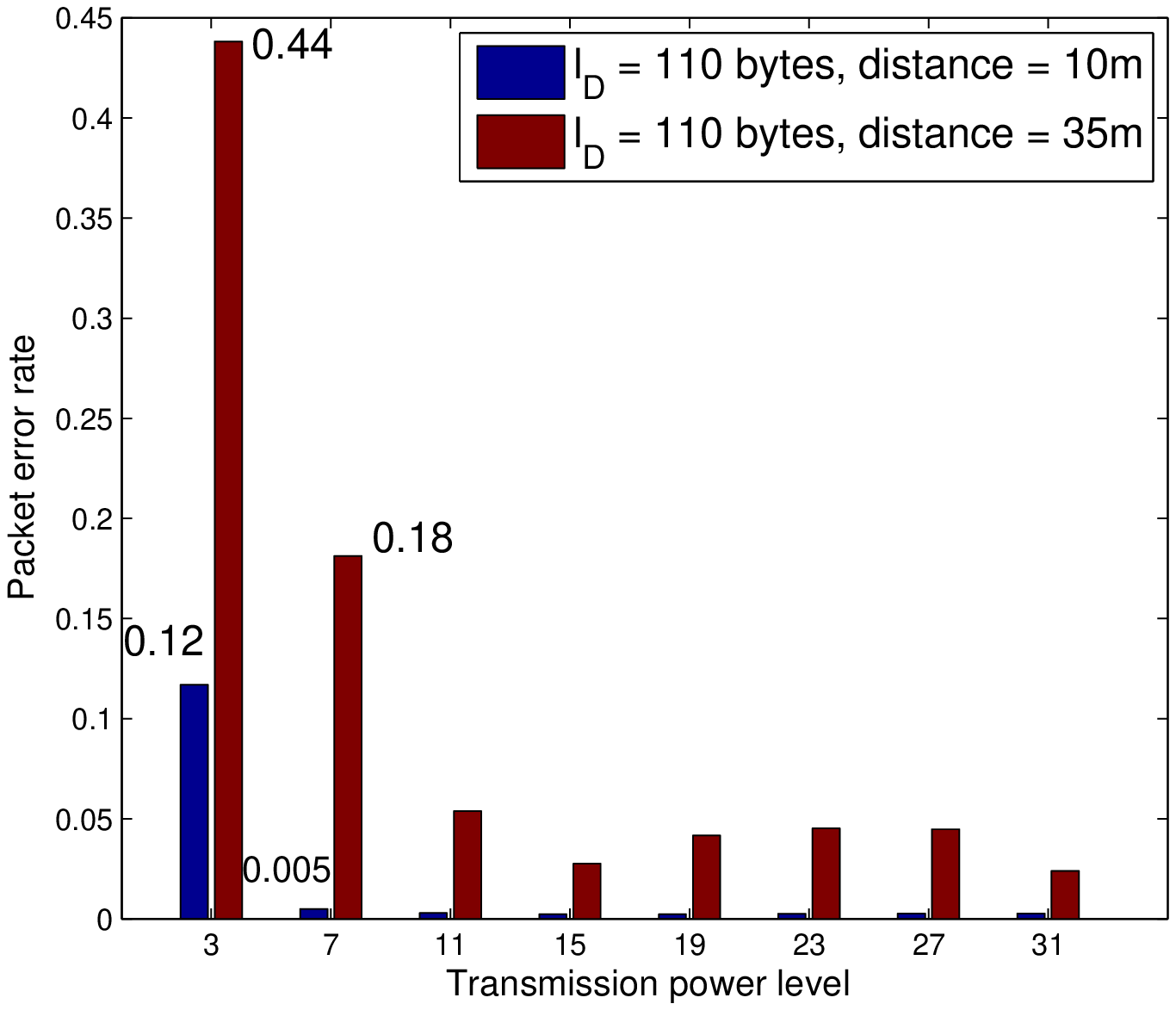}
\caption{PER for $l_D$=110 bytes}
\label{fig:per_d10_d35}
\end{minipage}%
\begin{minipage}[t]{0.5\columnwidth}
\centering
\includegraphics[width=2.0in, height=1.8in]{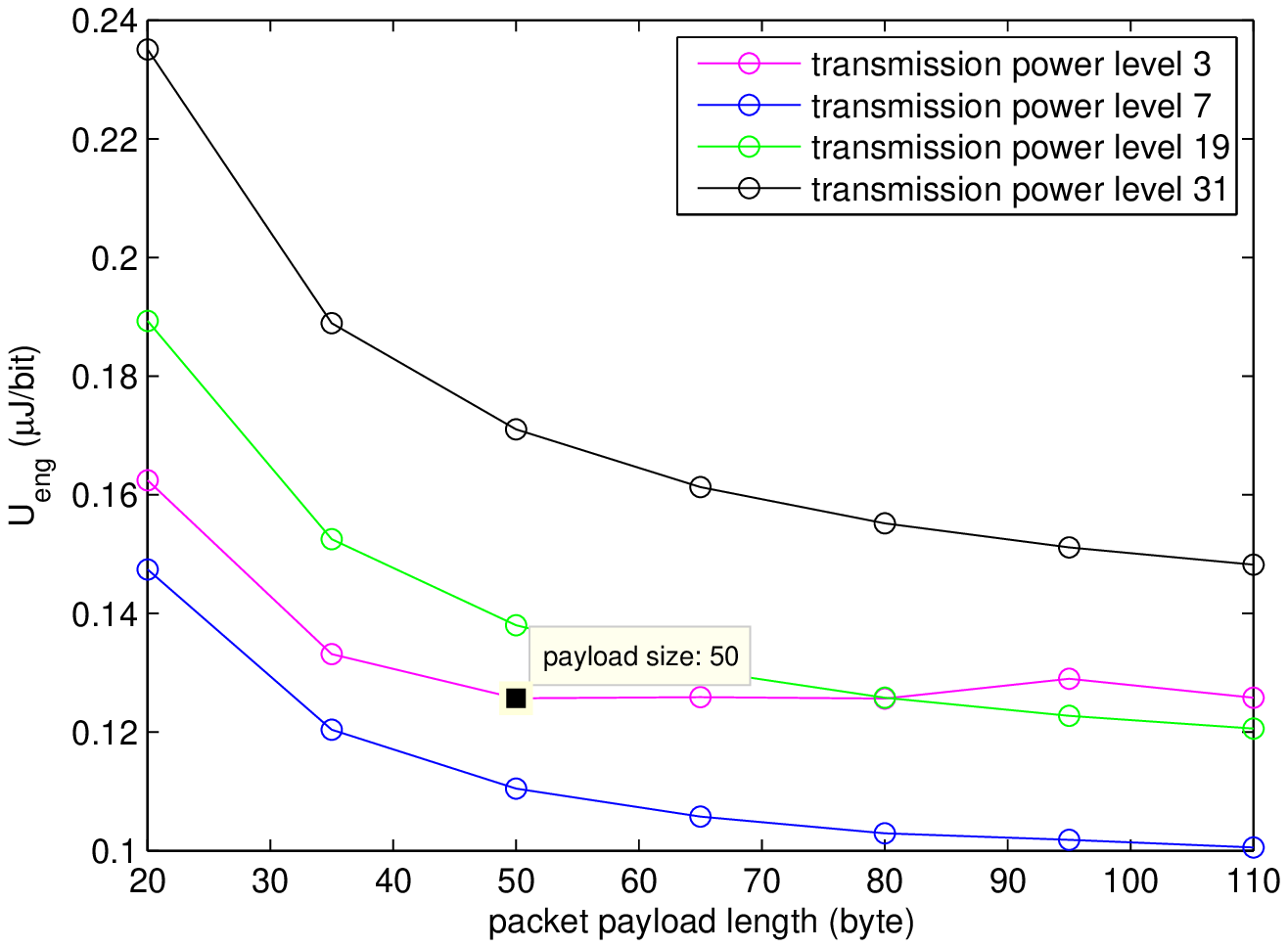}
\caption{$U_{eng}$ at 35m}
\label{fig:energy_vs_l}
\end{minipage}%
\end{figure}

\textbf{Reasoning.} For 10m, the increase of $U_{eng}$ is intuitively due to increased energy consumption with higher $P_{TX}$. However, for 35m, the same intuition does not fully apply. {\em This difference is resulted from the tremendous decrease of PER when the SNR crosses the border of the transitional zone and reaches the low-loss zone}. This is verified in~\figref{fig:per_d10_d35}, which plots the PER for each power level for $l_D = 110 \textrm{ bytes}$. It shows that, when $P_{TX}$ increases from 3 to 7, the link of 35m moves from the transitional zone to the low-loss zone (SNR from 5.5dB to 15.8dB), with PER decrease of approximately 0.26. On the other hand, the link for distance 10m is always in the low-loss zone from $P_{TX}=3$ where the decrease of PER does not cause significant reduction of $U_{eng}$.

To further investigate the optimal value of $l_D$ for energy consumption at different $P_{TX}$, we plot $U_{eng}$ against $l_D$ under different $P_{TX}$ for  35m in~\figref{fig:energy_vs_l}. It can be observed that the optimal $l_D$ is 50 bytes for $P_{TX}=3$ while it is the maximal $l_D$ (110 bytes in our case) for other $P_{TX}$. {\em The underlying reason for this difference is that when the link is in the high-impact zone of payload size on PER, increasing $l_D$ can cause higher PER thus higher $U_{eng}$}. ~\figref{PER3} shows that the PER increases by 0.12 from $l_D=20 \textrm{ bytes}$ to $l_D=110\textrm{ bytes}$ when $SNR=5.5dB$ at $P_{TX}=3$.

\textbf{Implication.} The above results imply that {\em the joint effect of $l_D$ and PER on energy efficiency can differ significantly when SNR is in different zones or moves from one zone to the other}. Specifically, to minimize $U_{eng}$, the following practical guidelines can be useful. (1) Increase $P_{TX}$ so that SNR moves to the low loss zone; if SNR is already in the low-loss zone, decrease $P_{TX}$ so that the link just barely stays in the low-loss zone. (2) Use maximal payload size to further minimize $U_{eng}$. (3) If the link can only stay in the transitional zone no matter what $P_{TX}$ is chosen, then a medium $l_D$ may be chosen to provide better energy efficiency.

%%%%%%%%%%%%%%%%%%%%%%%%%%%%%%%%%%%%%%%%%%%%%% Delay %%%%%%%%%%%%%%%%%%%%%%%%%%%%%%%%%%%%%%%%%%%%%%%%%%%%%%%%%%%

\subsection{Delay}
\label{sec:delay}

The delay perceived by a packet consists of two parts: queuing delay and service time delay. In our experiment, delay is measured for every received packet. To quantitatively answer how all the layer stack parameters contribute to the delay performance, we investigate the average delay against SNR for different traffic workloads under four typical MAC configurations: (a) no queue and no retransmission, (b) no queue but with retransmission, (c) with a queue and with retransmission and (d) with a larger queue and with retransmission. The maximum allowed queue size is determined by parameter $Q_{max}$. $Q_{max}=1$ means ``no queue'' (of other packets).

\begin{figure}[b!]
\centering
\includegraphics[width=1.0\columnwidth]{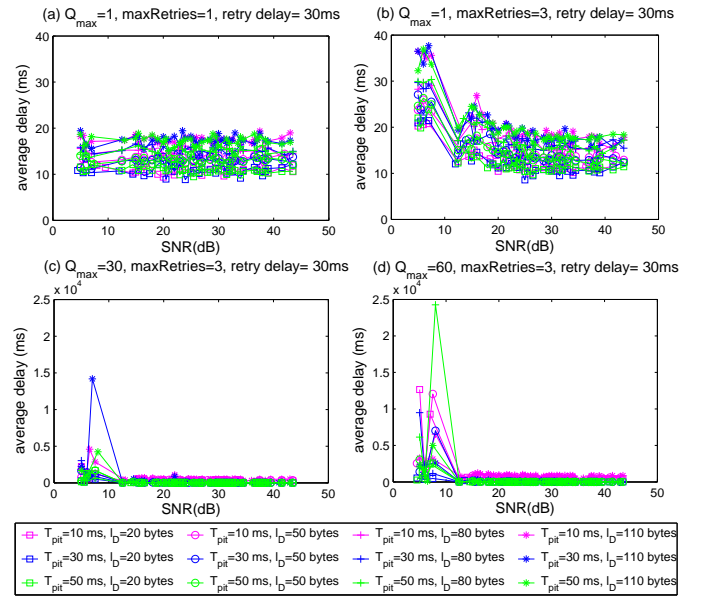}
\caption{Average delay under different parameter configurations}% change the name
\label{fig:delay1}
\end{figure}

\textbf{Observation.} The delay performance under the four MAC configurations is shown in~\figref{fig:delay1}(a),~\figref{fig:delay1}(b),~\figref{fig:delay1}(c) and~\figref{fig:delay1}(d) respectively. We highlight that distinct delay behaviors can be observed particularly when the link is in the transitional zone. Specifically, when $Q_{max}=1$, the delay in~\figref{fig:delay1}(a) and~\figref{fig:delay1}(b) has the same order of magnitude. When $Q_{max}=30 \textrm{ or } 60$, while the delay in~\figref{fig:delay1}(c) and~\figref{fig:delay1}(d) has the same order of magnitude, it is much larger than the delay with $Q_{max}=1$, when the link is in the transitional zone. This difference is due to queuing delay, as to be explained below. 

\textbf{Reasoning.} In order to understand the conditions for the occurrence of queuing, we first introduce the concept of system utilization and then develop an empirical model of average service time to quantify when and how queuing occurs depending on the stack parameter configurations.

Following queuing theory (see, e.g. \cite{Kleinrock75}), the system utilization, denoted by $\rho$, is defined to be 
\begin{equation} \label{equ:ro}
	{\rho} = \frac{\overline{T}_{service}}{T_{pit}}
\end{equation} 
where $T_{pit}$ denotes the packet inter-arrival time and $\overline{T}_{service}$ the average service time. From queuing theory, it is known that while the queuing delay does not change much with $\rho (< 1)$, it increases extremely quickly when $\rho \to 1$ and will not be bounded if $\rho > 1$ with no dropping mechanism employed. 

In our study, $T_{pit}$ is a configurable parameter and was fixed for each experiment run. To find $\overline{T}_{service}$, we introduce an empirical model. Under this model, the system is treated as a blackbox for each packet. As such, the service time depends on (1) $\mathit{T_{SPI}}$ -- the one-time hardware SPI bus interface loading time of a data frame; (2) $\mathit{T_{frame}}$ -- the time to transmit a frame consisting of data packet payload $l_D$ and 17 bytes overhead, i.e. 11 bytes MAC header plus 6 bytes PHY header; (3)  $\mathit{T_{MAC}}$ -- MAC layer delay consisting of two parts: $T_{TR}$ and $T_{BO}$, where $T_{TR}$ is the turn around time and $T_{BO}$ is the average value of initial backoff period with $T_{TR}=0.224ms$ and $T_{BO}=5.28ms$ according to the TinyOS radio stack; (4) $\mathit{T_{ACK}}$ -- the ACK frame transmission time if ACK frame is received, and based on prior tests $T_{ACK}\approx 1.96 ms$ for our study; (5) $\mathit{T_{waitACK}}$ -- the maximum waiting period for ACK frames (In the TinyOS implementation, the maximum software ACK waiting period is $T_{waitACK}=8.192ms$); (6) $\mathit{\overline{N}_{retry}}$ -- the number of (re-)transmissions to deliver the packet successfully; (7) $\mathit{D_{retry}}$ -- the retry delay, i.e. the time between two consecutive retransmissions. Specifically, there are two cases depending on if the packet is successfully transmitted. Accordingly, 

\begin{itemize}
	\item If $\overline{N}_{retry}\leq N_{maxRetries}$,  
	\begin{equation}\label{avg_servicetime1}
		\overline{T}_{service} =T_{SPI}+T_{succ}+(\overline{N}_{retry}-1)\cdot T_{retry}
	\end{equation}
	\item If $\overline{N}_{retry}> N_{maxRetries}$, 
		\begin{equation}\label{avg_servicetime2}
		\overline{T}_{service}=T_{SPI}+T_{fail}+(N_{maxRetries}-1)\cdot T_{retry}
	\end{equation}
\end{itemize}
where
\begin{eqnarray*}
T_{succ} &=& T_{MAC}+T_{frame}+ T_{ACK}\\
T_{fail}&=& T_{MAC}+T_{frame}+ T_{waitACK}\\
T_{retry} &=& D_{retry}+T_{MAC}+T_{frame}+T_{waitACK}
\end{eqnarray*}

\begin{figure}[ht!]
\centering
\includegraphics[width=0.8\columnwidth]{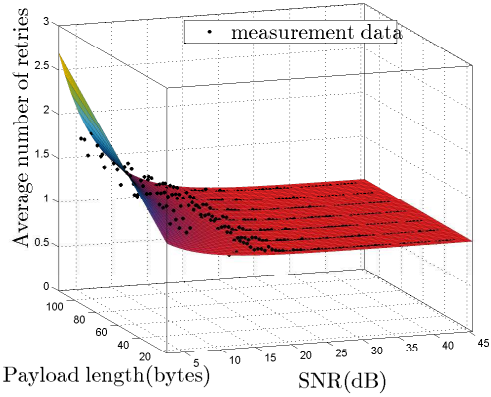}
\caption{Modeling average number of transmissions}% change the name
\label{avgNretry}
\end{figure}

To decide ${\overline{N}_{retry}}$ in the service time model, an empirical method is applied. Based on the measured average number of transmissions with respect to $l_D$ and SNR, as illustrated in ~\figref{avgNretry}, we first model its relation with SNR and then model its relation with $l_D$ and obtain the following exponential function:
\begin{equation} \label{avg_retry}
	\overline{N}_{retry}=1+\alpha\cdot l_D\cdot \exp(\beta\cdot \textrm{SNR}),
\end{equation}
where $\alpha$ and $\beta$ are the model coefficients. Applying curving fitting, we obtain $\alpha=0.02$ and $\beta=-0.18$ which are significant at $95\%$ confidence level. 

With ${\overline{N}_{retry}}$, the empirical model of average service time, i.e. (\ref{avg_servicetime1}) and (\ref{avg_servicetime2}), can be readily used. In particular, Table~\ref{tab:systemUti} lists the corresponding service time and system utilization for (part of) the parameter configurations considered in~\figref{fig:delay1}. 

\begin{table}[htbp]
	\centering
	\caption{System utilization}
		\begin{tabular}{|c|c|c|c|c||c|}
		\hline
		\textbf{${T_{pit}}$}(ms) & SNR(dB) & \textbf{$l_D$} & maxRetries & \textbf{$T_{service}$}(ms) & \textbf{$\rho$}\\ 
		\hline
		\hline
		30 & 10 & 110 & 1 & 24.12 & 0.804\\ 
		%\hline
		%30 & 20 & 110 & 1 & 24.12 & 0.804\\ 		
		\hline
		30 & 10 & 110 & 3 & 37.08 & 1.236\\
		\hline
		30 & 20 & 110 & 3 & 21.39 & 0.713\\
		\hline
		\end{tabular}
	\label{tab:systemUti}
\end{table}

Specifically, Table~\ref{tab:systemUti} shows two under-utilized cases (where  $\rho <1$) and one over-utilized case (where  $\rho >1$). For the over-utilized case, an excessive delay can be expected from queuing theory results, explaining the observation from~\figref{fig:delay1}, where that the finite rather than unbounded maximal delay is caused by limited buffer space as well as a limited number of retries.

\textbf{Implication.} To control the delay, the system should be configured with a proper system utilization level $\rho$. When $\rho< 1$, there is little or no queuing or queuing delay. However, when $\rho\geq1$, a significant delay may be encountered. In addition, the larger the maximal queue size, the larger the queuing delay. Since $\rho$ is jointly determined by $l_D$, $T_{pit}$, $maxRetries$, retry delay and SNR, more specifically, $P_{TX}$ and distance, it is necessary to take them all into consideration to minimize the delay. Specifically, there are two practical guidelines helpful to avoid queuing delay. One is to completely remove the queue at the MAC layer. Another is, for the given $l_D$, maxRetries, retry delay and SNR, to choose the packet inter-arrival time properly so that the system utilization is smaller than 1. In addition, if queuing delay is avoided or reduced, reducing $l_D$ and $maxRetries$ or increasing $P_{TX}$ can reduce the service time delay therefore further reducing the overall delay.

%%%%%%%%%%%%%%%%%%%%%%%%%%%%%%%%%%%%%%%%%%%%%% Goodput %%%%%%%%%%%%%%%%%%%%%%%%%%%%%%%%%%%%%%%%%%%%%%%%%%%%%%%%%%%

\subsection{Goodput}
\label{sec:goodput}

Goodput is the application level throughput, i.e., the number of useful information bits received at the receiving node per unit of time. In our experiment, the goodput is calculated for each experiment run by using the total number of information bits received during this run divided by the time between receiving the first packet and the last packet. Similar to that for delay, we investigate goodput against SNR for different traffic workloads also under four typical MAC configurations: (a) no queue and no retransmission, (b) no queue but with retransmission, (c) with a queue but no retransmission and (d) with a queue and with retransmission.  

\textbf{Observation.} The goodput performance under the four MAC configurations is shown in~\figref{throughput}(a), \figref{throughput}(b), \figref{throughput}(c) and~\figref{throughput}(d) respectively. It can be observed that {\em for all parameter combinations, when the link is in the transitional zone, the goodput is lower than the goodput when the link is in low-loss zone}. In addition, it is also interesting to observe that the goodput for the same $T_{pit}$ and $l_D$ behaves similarly under different MAC configurations except for $T_{pit}$ of 10ms. When $T_{pit}=10ms$, goodput increases if the queue is used at the MAC layer. It is also noticed that MAC retransmission does not have much impact on the goodput. 

\begin{figure}[t!]
\centering
\includegraphics[width=1.0\columnwidth, height=2.5in]{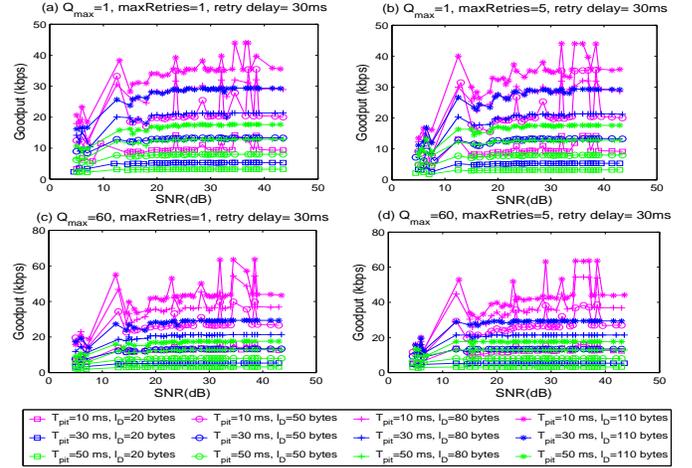}
\caption{Goodput under different parameter configurations}
\label{throughput}
\end{figure}

\nop{
\begin{figure*}
\begin{minipage}[t]{1.0\columnwidth}
\centering
\includegraphics[width=1.0\columnwidth, height=2.5in]{fig_throughput_subplot.eps}
\caption{Goodput under different parameter configurations}
\label{throughput}
\end{minipage}%
\begin{minipage}[t]{1.0\columnwidth}
\centering
\includegraphics[width=1.0\columnwidth, height=2.2in]{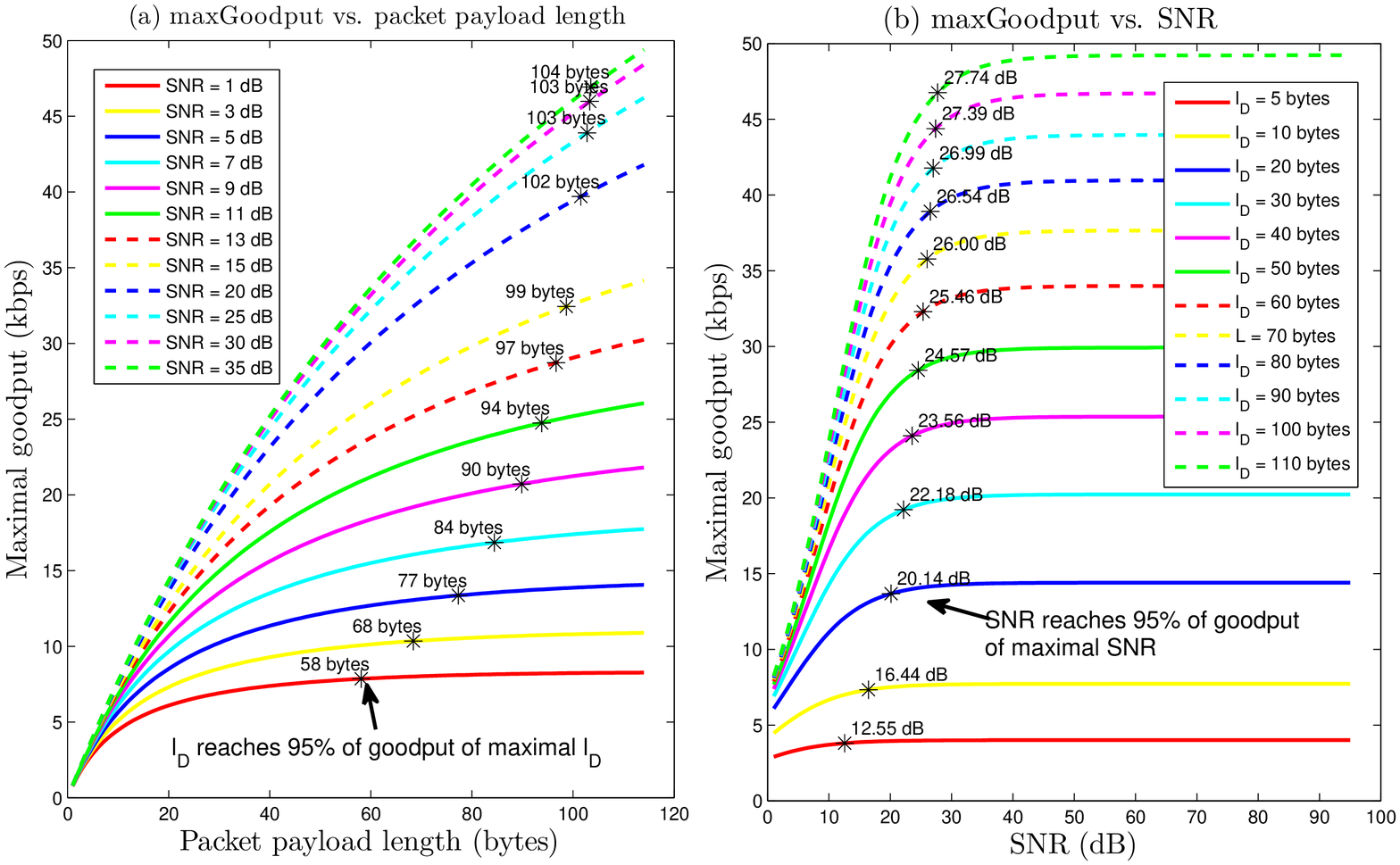}
\caption{Maximal goodput}
\label{fig:maxGoodput}
\end{minipage}%
\end{figure*}
}

\textbf{Reasoning.} To understand the depicted goodput behaviors, we first discuss about the goodput under the assumption of no radio loss during transmission. According to the definition, the goodput can be computed by the received $l_D$ divided by the packet inter-arrival time at the receiving node $T_{receiver}$. If the system utilization $\rho<1$ and there is no packet loss during transmission, $T_{receiver}$ equals the packet inter-arrival time $T_{pit}$ at the sending node. When $\rho\geq 1$ and the queue is big enough, all the packets are queued for transmission and then $T_{receiver}$ is equal to $T_{service}$ at the sending node (due to fixed packet length for each experiment run). Therefore, the goodput considering no packet loss can be expressed as:% follows. 
\begin{equation} \label{equ:goodput}
	\mathrm{goodput} = \frac{{l_D}}{T_{receiver}} =\frac{{l_D}}{max(T_{pit}, T_{service})}
\end{equation} 

Using (\ref{equ:goodput}), we investigate the impacts of different stack parameters on the goodput. If the system utilization $\rho<1$, decreasing $T_{pit}$ can increase the goodput until $T_{pit}$ close to $T_{service}$ where goodput is maximized to $l_D/T_{service}$ in the low-loss zone. After that, continue reducing $T_{pit}$ can cause $\rho\geq 1$, resulting in queuing loss and a large queue can reduce the chance of queue overflow. This explains the observation from~\figref{throughput}. Specifically, when $T_{pit}$ decreases from 50ms to 30ms and then to 10ms, the goodput of the same $l_D$ increases. For the case of $T_{pit}=10ms$, it can be found from the introduced empirical service time model that $\rho\geq1$. In this case, a large allowed queue size will reduce packet loss and hence improve the goodput. For the small impact difference of MAC retransmission, we infer that this is because of the opposite effect of retransmission on the radio loss and queuing loss and we leave more discussion on this issue to Sec. \ref{sec:plr}.

\textbf{Implication.} As discussed above, (\ref{equ:goodput}) implies the maxmial goodput (maxGoodput) to be $\mathrm{maxGoodput} = \frac{{l_D}}{T_{service}}$. In other words, maxGoodput depends on both $l_D$ and $T_{service}$. The latter is further affected by various parameters particularly transmission power as implied by (\ref{avg_servicetime1}), (\ref{avg_servicetime2}) and (\ref{avg_retry}). To investigate the impact of $l_D$ and $P_{TX}$ on maxGoodput, ~\figref{fig:maxGoodput} plots the maxGoodput performance against $l_D$ and SNR. It can be observed a general trend: bigger $l_D$ and larger SNR always lead to higher maxGoodput. However and more appealingly, the figure also shows that {\em using a moderate transmission power level or a moderate payload size is able to reach an excellent level for maxGoodput}, e.g., 95\% of that with maximal $l_D$ as marked in~\figref{fig:maxGoodput}(a) or 95\% of that with maximal $P_{TX}$ as marked in~\figref{fig:maxGoodput}(b) respectively. This brings new insights to choose optimal values for $l_D$ and/or $P_{TX}$ when PLR, energy consumption and goodput are considered together.

\begin{figure}[t!]
\centering
\includegraphics[width=1.0\columnwidth, height=2.2in]{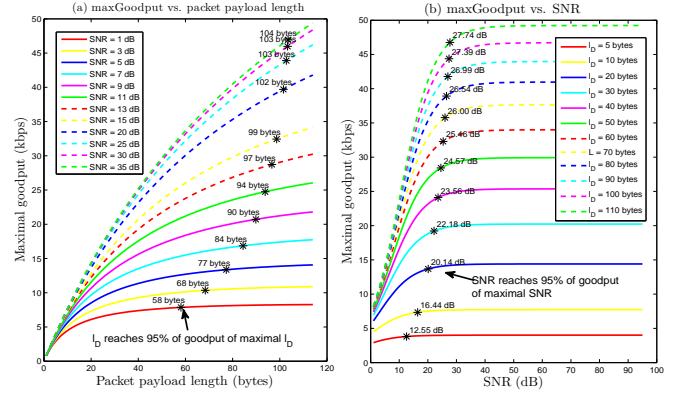}
\caption{Maximal goodput}
\label{fig:maxGoodput}
\end{figure}

Practically, to maximize goodput, the key is to keep system busy or queuing for transmission and at the same time try to minimize the service time. Specifically, increasing $P_{TX}$ can effectively reduce the service time $T_{service}$. Despite that larger payload size results in longer $T_{service}$, the goodput still increases with $l_D$. Therefore, selecting maximal $P_{TX}$ and maximal $l_D$ can result in the best goodput. In addition, $T_{pit}$ may be configured such that the system utilization $\rho\geq 1$. Furthermore, maintaining a large queue at the MAC will also help to achieve a high goodput level. Nevertheless, if it is also required to limit the energy consumption and/or PLR, more of the above investigation results may be employed and a moderate transmission power and/or a moderate payload size may suffice.

%%%%%%%%%%%%%%%%%%%%%%%%%%%%%%%%%%%%%%%%%%%%%% Packet loss rate %%%%%%%%%%%%%%%%%%%%%%%%%%%%%%%%%%%%%%%%%%%%%%%%%%%%%%%%%%%

\subsection{Packet Loss Rate}
\label{sec:plr}

Packet loss rate (PLR) consists of two parts: (1) $PLR_{queue}$ -- queuing loss rate, i.e. the ratio of packet loss due to buffer overflow, and (2) $PLR_{radio}$ -- radio loss rate, which refers to the ratio of packets lost (or indeed not correctly received) on radio transmission. Again, four typical MAC configurations are considered, namely (a) no queue and no retransmission, (b) no queue but with retransmission, (c) with a queue but no retransmission and (d) with a queue and with retransmission. 

\begin{figure}[t!]
\centering
\includegraphics[width=1.0\columnwidth]{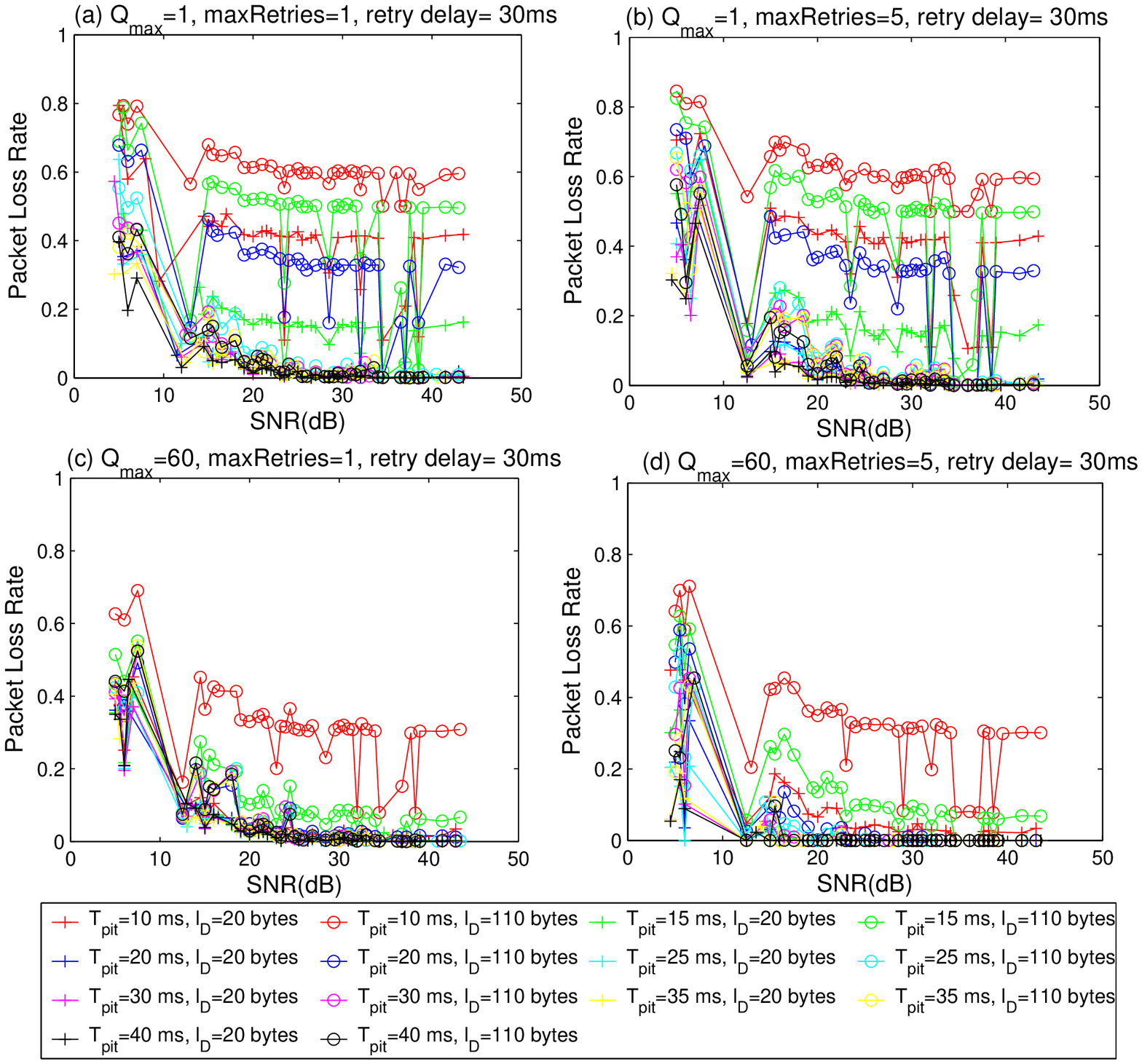}
\caption{Packet loss rate under different parameter configurations}% change the name
\label{fig:PLR}
\end{figure}

\textbf{Observation.} The measured PLR performance under the four MAC configurations is shown in~\figref{fig:PLR}(a), ~\figref{fig:PLR}(b), ~\figref{fig:PLR}(c) and ~\figref{fig:PLR}(d) respectively. It can be observed that PLR is higher in the transitional zone and decreases with SNR. In addition, queue size and retransmission are effective to reduce PLR in the transitional zone of the link. 

In order to provide more in-depth understanding of PLR, the corresponding $PLR_{queue}$ and $PLR_{radio}$ of~\figref{fig:PLR} are plotted in~\figref{fig:qloss} and~\figref{fig:radioloss} respectively.

\begin{figure}[t!]
\centering
\includegraphics[width=1.0\columnwidth, height=2.2in]{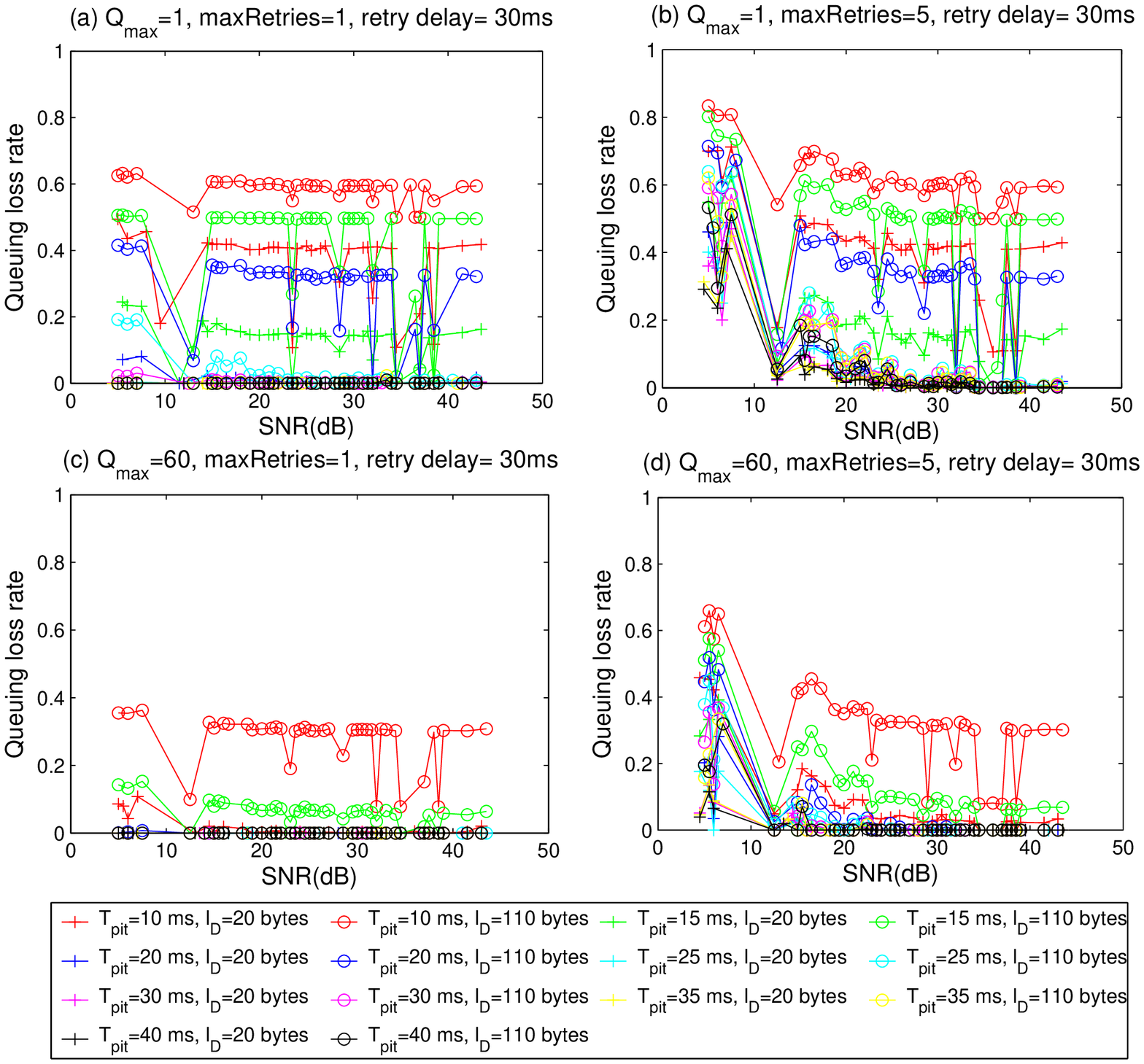}
\caption{Queuing loss under different parameter configurations}
\label{fig:qloss}
\end{figure}

\begin{figure}[t!]
\centering
\includegraphics[width=1.0\columnwidth, height=2.2in]{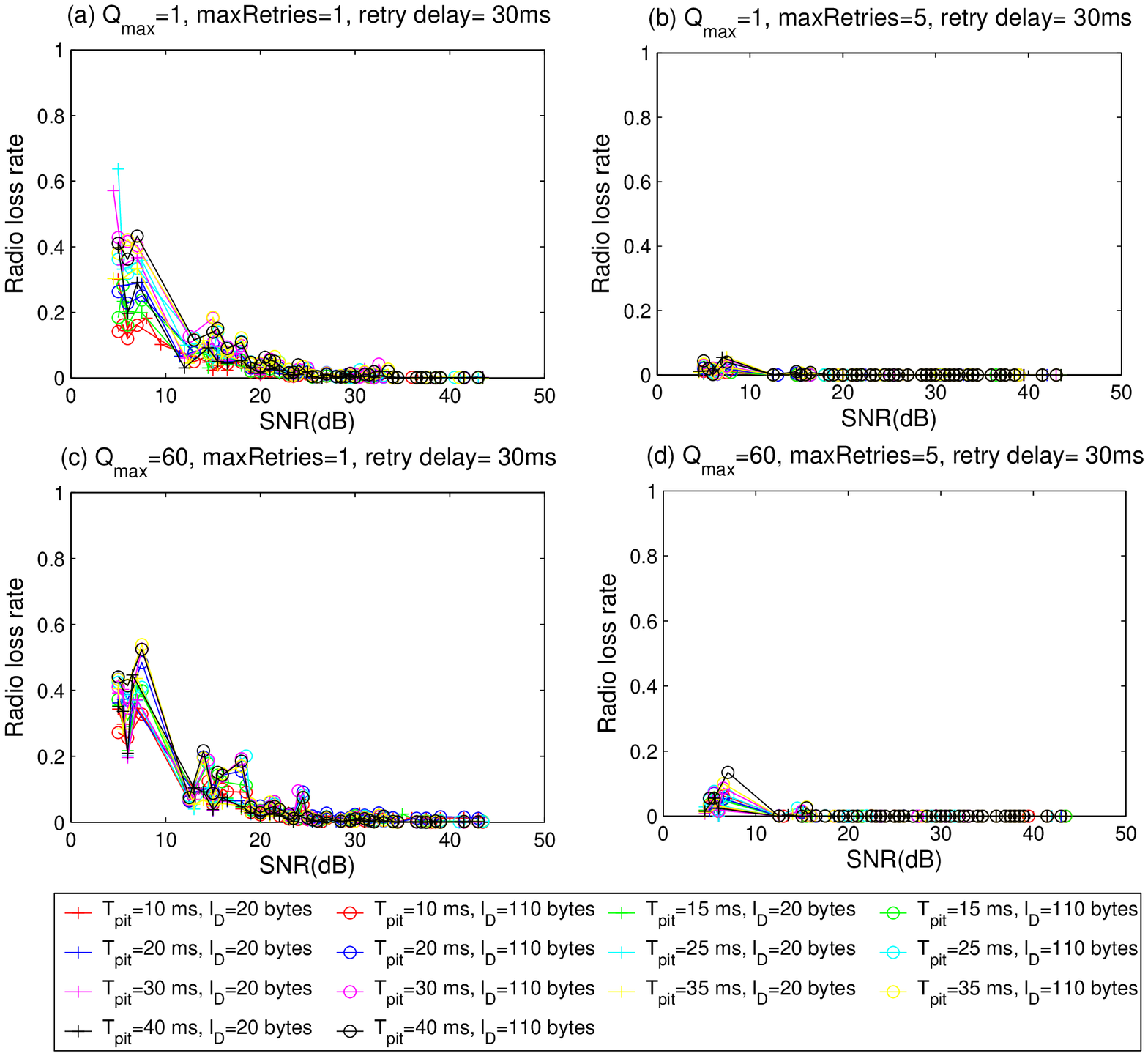}
\caption{Radio loss under different parameter configurations}
\label{fig:radioloss}
\end{figure}

\nop{
\begin{figure*}
\begin{minipage}[t]{1.0\columnwidth}
\centering
\includegraphics[width=1.0\columnwidth, height=2.2in]{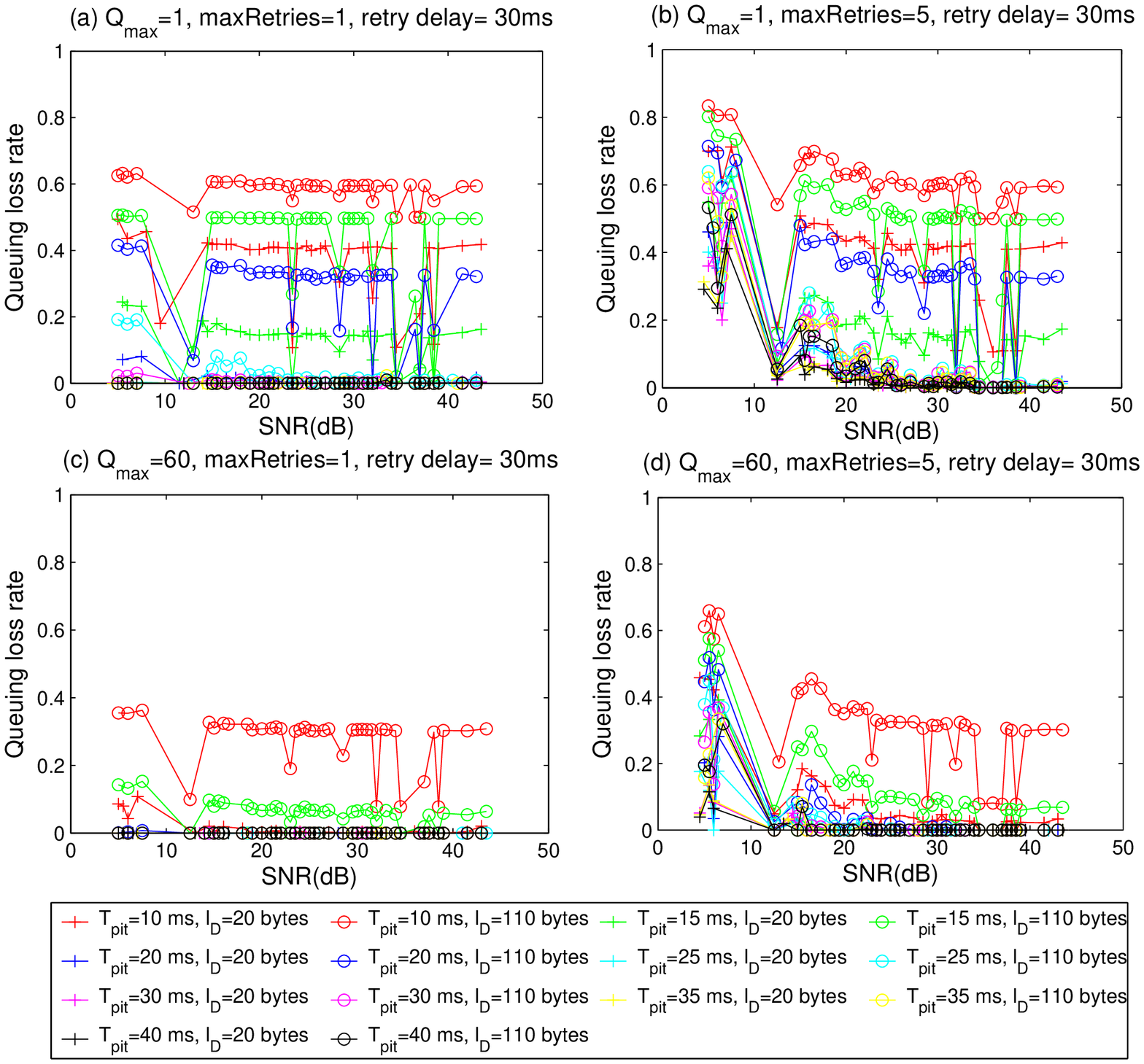}
\caption{Queuing loss under different parameter configurations}
\label{fig:qloss}
\end{minipage}%
\begin{minipage}[t]{1.0\columnwidth}
\centering
\includegraphics[width=1.0\columnwidth, height=2.2in]{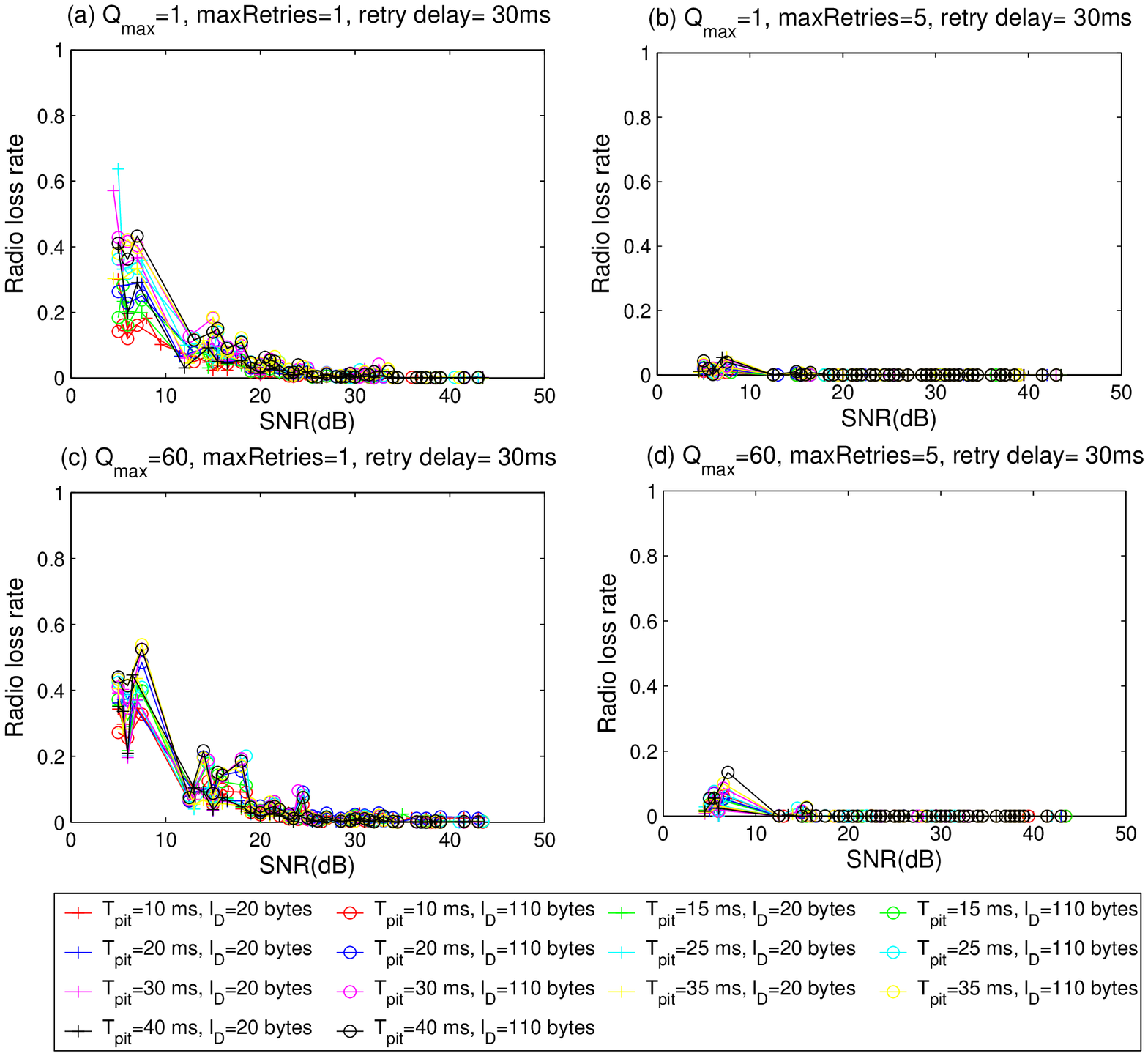}
\caption{Radio loss under different parameter configurations}
\label{fig:radioloss}
\end{minipage}%
\end{figure*}
}

\begin{figure}[t!]
\centering
\includegraphics[width=0.6\columnwidth]{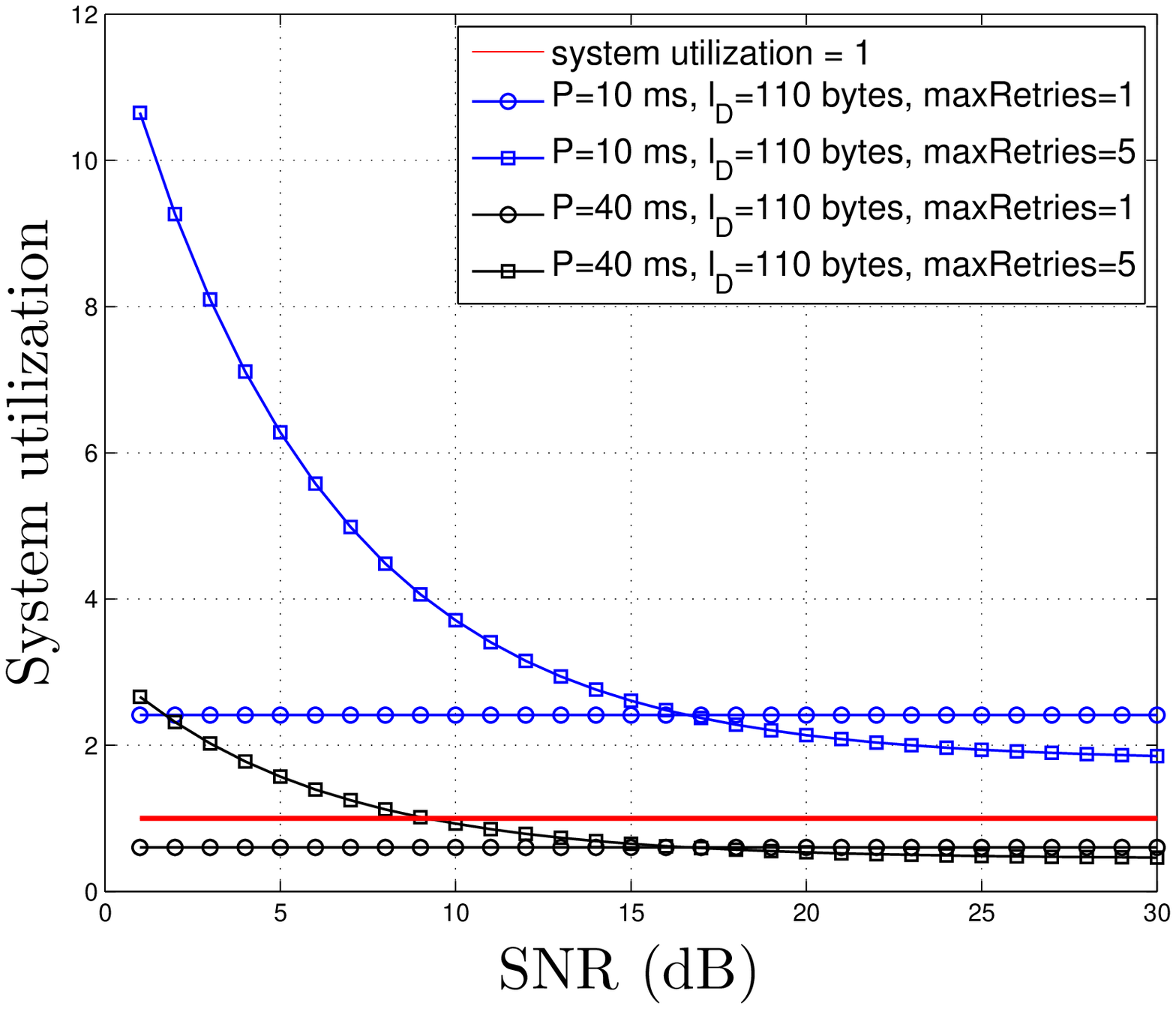}
\caption{System utilization}% change the name
\label{fig:sysUtility}
\end{figure}

\textbf{Reasoning.} For queuing loss, the system utilization $\rho$ is an effective indicator, e.g., when $\rho >1$, the excess part has to be dropped. To demonstrate this,~\figref{fig:sysUtility} plots $\rho$ against SNR under different combinations of $T_{pit}$, $l_D$ and maxRetries, partly in correspondence to~\figref{fig:PLR}. The figure shows two typical causes of queuing: (1) When small $T_{pit}$ and large $l_D$ are used for transmission, the system utilization $\rho$ can easily exceed 1 no matter what SNR or MAC setting is. For example, the system utilization at $T_{pit}=10 \textrm{ ms}$ and $l_D=110 \textrm{ bytes}$ is bigger than 1. It explains the queuing loss behaviors in~\figref{fig:qloss} (a); (2) When the link is in the transitional zone with retransmission allowed, $T_{service}$ will increase making $\rho\geq 1$. For example, the system utilization at $T_{pit}$ 40ms and $l_D=110 \textrm{ bytes}$ turns larger than 1 after SNR of 10dB. This explains the similar behaviors observed in ~\figref{fig:qloss} (b). Corresponding to~\figref{fig:qloss} (a) and (b),~\figref{fig:qloss} (c) and (d) indicate that a larger queue size can reduce the queuing loss. This is due to that it is more difficult to overflow with a larger size queue.

For radio loss,~\figref{fig:radioloss} plots $PLR_{radio}$ under different parameter configurations. As expected and the purpose of retransmission, the figure illustrates that the radio loss ratio is significantly reduced if retransmission is allowed. However, it is worth highlighting that retransmission can cause an increase on the system service time as implied by (\ref{avg_servicetime1}) and (\ref{avg_servicetime2}). Consequently, the system utilization $\rho$ will also increase. An overly large maxRetries may even lead to $\rho >1$, causing big queuing loss. This explains why~\figref{fig:radioloss}(b)/(d) shows lower radio loss than~\figref{fig:radioloss}(a)/(c), while their corresponding figures for queuing loss in~\figref{fig:qloss} show the opposite.

\textbf{Implication.} The above results imply the following parameter optimization technique to minimize PLR. Specifically, increasing $P_{TX}$ so that the link stays in the low-loss zone is effective to reduce both queuing loss and radio loss. MAC retransmission is helpful to reduce radio loss but it could worsen queuing loss. A reasonable value of maxRetries is hence necessary. For example, for our considered indoor environment, this value can be obtained from the empirical model of average retries (\ref{avg_retry}). After these, $T_{pit}$ and $l_D$ should then be properly chosen to keep system utilization $\rho<1$ to avoid queuing loss. If $\rho\geq 1$, a larger queue is recommended to reduce queuing loss.

\section{Discussion}\label{sec:dis}

\subsection{Consideration on Parameter Configuration under Different Link Quality Conditions}

In Sec.~\ref{sec:radio-channel} and Sec.~\ref{sec:analysis}, we have presented extensive results of the experimental study and reasoned the joint impacts of various parameters at multiple layers on the aforementioned performance metrics. A crucial finding is that there exist different zones with distinct properties for PER, which is highly related to and affects other metrics, under different link quality (characterized by SNR) conditions or under different payload sizes. {\em A consequence of the existence of multiple zones is that an observation or result for one zone cannot and should not be generalized for the other zones}. 

Here, we highlight three example cases taken from the earlier discussion where the same parameter may have highly different impacts in different zones. One is that, while larger $L_D$ always leads to higher energy efficiency in the low-loss zone, it may cause extra retransmission due to higher PER caused by the larger size and thus lower energy efficiency in the transitional zone. In the second case, consider delay as the targeted QoS performance metric. To achieve a low delay, the system should be properly configured. In particular, when the link is in the traditional zone, if $maxRetries$ and $L_D$ are not carefully configured, they may increase $T_{service} $ and cause $\rho > 1$, resulting in significant delay. However, the same parameter configuration may only have limited impact on delay in the low-loss zone. In the third case, suppose goodput is the performance metric of interest. While increasing $L_D$ is effective to improve the goodput in the low-loss zone, it is not in the transitional zone due to higher PER.

\subsection{Trade-offs in Choosing Parameter Values}
\label{sec:tradeoff}

Our analysis on the joint parameter configuration impact has so far been conducted for each individual performance metric. It is worth highlighting that while one parameter configuration may achieve better or optimal performance for one metric, it may not do so for another metric or when multiple performance metrics need to be considered together. In fact, there are trade-offs when deciding the parameter values for different performance metrics. It is hence crucial to understand these tradeoffs and adopt proper parameter configurations to achieve multiple performance objectives. Here, we discuss briefly\nop{ due to space limitation,} how parameter configuration decisions may lead to performance optimization tradeoffs. Table \ref{tab:tradeoff} summarizes such tradeoffs.

In general, the system utilization $\rho$ and the queue configuration control the QoS related performance and their trade-offs. Larger queue size and higher number of allowed retransmissions can reduce PLR and increase goodput. However, they will result in an increased delay. If the link is in the transitional zone, increasing transmission power such that the SNR moves to the low-loss zone improves all QoS metrics and even the energy consumption. If the link can only stay in the transitional zone, the QoS metrics and energy efficiency are more sensitive to the stack parameter configuration. In particular, a moderate packet payload size may better balance the trade-offs between the performance metrics. 
\begin{table}[ht!]
	\centering
%%	\caption{Stack parameter impact on performance trade-offs \\(P, N and - represents positive, negative and no impacts respectively)}
	\caption{Stack parameter impact on performance trade-offs ($P$, $N$ and $-$: positive, negative and no impact respectively)}
		\begin{tabular}{|l|p{2cm}|c|c|c|c|}
		\hline
		Layer & Parameters with increasing value & Delay & Goodput & PLR & $U_{eng}$\\ 
		\hline
		\hline
		APP & $T_{pit}$ when $\rho<1$ & - & N & - & - \\ \cline{2-6}
		& $T_{pit}$ when $\rho\geq 1$ & P & - & P & - \\ \cline{2-6}
		& payload size & N & P & N & P/N \\
		\hline
		MAC	& queue size and retransmission & N & P & P & - \\
		%\hline
	  %& Maximal number of retires(maxRetries) & & & &  \\
		%\hline
		%& Retry delay(ms) & & & & \\
		\hline
		PHY	& $P_{TX}$ & P & P & P & P/N \\
		%\hline
		\hline
		\end{tabular}
	
	\label{tab:tradeoff}
\end{table}

\section{Conclusion} \label{sec:conclusion}

This paper has presented an extensive experimental study to understand the performance of 802.15.4 wireless link for WSNs. Based on the measurement data, in addition to analyzing the joint parameter impacts on energy efficiency and QoS-related performance metrics, we have revealed several interesting findings, based on which practical guidelines for parameter configurations were suggested. These findings, importantly, indicate that we should have different considerations in response to different link quality zones and on multi-parameter optimization decisions. These findings are helpful to avoid incorrect generalization on parameter optimization: An optimal parameter configuration for one performance metric under one link quality condition may degrade the performances of other metrics and/or under other link quality conditions. In addition, we have introduced several empirical models (i.e., $\overline{N}_{retry}$ and $\overline{T}_{service}$) that can help for performance analysis and reason the results. Finally, we are convinced with the necessity of conducting such an extensive experimental study and believe that our results are valuable to the community for obtaining better understanding on wireless link performance in WSNs and specifically about how multi-layer parameters may be jointly configured to fulfill QoS requirements of applications.

%\bibliographystyle{abbrv}
%\bibliography{bibtex/yanzhangbib}
%\bibliography{bibtex/infocomm2015}

\end{document}